\documentclass[usenatbib,referee]{mn2e}
\bibpunct{(}{)}{;}{a}{}{,}
\usepackage{ram}
\usepackage{graphicx}
\usepackage{lscape}
\usepackage{txfonts}
\usepackage{float}
\usepackage[]{caption}
\usepackage{epstopdf}
\def\ha{H$\alpha$}
\def\hb{H$\beta$}

\def\othree{[O\,{\sc{iii}}]}

\def\ntwo{[N\,{\sc{ii}}]}
\def\stwo{[S {\sc{ii}}]}
\def\o{[O\,{\sc{i}}]}
\def\ca2T{Ca\,{\sc{ii}} Triplet}
\def\HI{H\,{\sc i}}
\def\kms{km\,s$^{-1}$}

\begin{document}

\title{AGN Activity and Black Hole Masses in Low Surface Brightness Galaxies}

\author[S. Ramya et al.]{S. Ramya$^{1}$\thanks{E-mail : ramya@iiap.res.in (SR)},
T. P. Prabhu$^{1}$\thanks{E-mail : tpp@iiap.res.in (TPP)}, 
M. Das$^{1, 2}$\thanks{E-mail : mousumi@iiap.res.in (MD)}\\
$^{1}$ Indian Institute of Astrophysics, Koramanagala, Bangalore-34, India. \\
$^{2}$ Birla Institute of Technology and Science - Pilani, Hyderabad Campus, Jawahar Nagar, Shameerpet
Mandal, Hyderabad, 500078, India. }

\date{Received / Accepted}

\pagerange{\pageref{firstpage}--\pageref{lastpage}} \pubyear{2006}

\maketitle

\begin{abstract}
We present medium resolution optical spectroscopy of a sample of nine Low Surface Brightness (LSB) galaxies.
For those that show clear signatures of AGN emission, we have disentangled the AGN component from 
stellar light and any Fe\,{\sc i} and Fe\,{\sc ii} contribution.
We have decomposed the \ha \ line into narrow and broad components and determined the velocities of the 
broad components; typical values lie between 900--2500 \kms. Of the galaxies in our study, UGC 6614, 
UGC 1922, UGC 6968 and LSBC~F568-6 (Malin~2) show clear signatures of AGN activity. We have calculated the 
approximate black hole masses for these galaxies from the \ha \ line emission using the virial approximation. 
The black hole masses are $\sim3~\times10^{5}~M_{\odot}$ for three galaxies and lie in the intermediate
mass black holes domain rather than the supermassive range. UGC 6614 harbors a BH of mass $3.8~\times10^{6}~M_{\odot}$; 
it also shows an interesting feature blueward of \ha \ and \hb \ implying outflow of gas or a one-sided jet streaming towards
 us. We have also measured the bulge stellar velocity dispersions using the  \ca2T lines and plotted these galaxies on the
 $M-\sigma$ plot. We find that all the three galaxies UGC 6614, UGC 6968 and F568-6 lie below the $M-\sigma$ relation for 
nearby galaxies. Thus we find that although the bulges of LSB galaxies may be well evolved, their nuclear black hole 
masses are lower than those found in bright galaxies and lie offset from the $M-\sigma$ correlation.
\end{abstract}


\section{\sc\bf Introduction}

Low Surface Brightness (LSB) galaxies are
an extreme class of late type spiral galaxies \citep{ImpeyBothun97}. They are poor in star  
formation (\citealt{Boissier.etal.2008}; \citealt{O'Neil.etal.2004}), have low metallicities \citep{mcgaugh94} 
and low dust masses 
(\citealt{Rahman.etal.2007}; \citealt{Hinz.etal.2007}). However, they are gas rich (\HI) and  
their \HI \ disks are far more extended than their diffuse stellar disks \citep{O'Neil.etal.2004}. Their disks 
have weak spiral arms and their bar perturbations are not as prominent as those seen in bright galaxies. 
This low level of disk activity can be attributed to the presence of dark matter halos \citep{deblok.etal.2001} that 
tend to prevent the formation of disk instabilities 
(\citealt{Mayer.Wadsley.2004}; \citealt{Mihos.etal.1997}; \citealt{galaz.etal.2011}). 
Surveys show that although LSB galaxies are common in our local universe they are preferentially found in low
 density environments \citep{rosenbaum.etal.2009} and can have widely varying morphologies;
starting from the populous dwarf LSB galaxies \citep{schombert.etal.2001} to the giant LSB galaxies (GLSB), of 
which Malin~1 is a good example \citep{pickering.etal.1997}.

Although a lot of work has been done towards understanding the stellar and gas content of LSB galaxies,
not much is known about their nuclear activity or central black hole (BH) masses. A significant fraction of GLSB 
galaxies have active galactic nuclei (AGN) that are often associated with massive bulges (\citealt{schombert98}; 
\citealt{impey.etal.2001}). 
The AGN may be radio bright \citep{das.etal.radio.2009} and visible at X-ray wavelengths as well 
(\citealt{naik.etal.2010}; \citealt{das.etal.xray.2009}). It is now well established that the black hole masses
in galaxies are correlated with their bulge velocity dispersions ($M-\sigma$) and bulge luminosities ($M-L$)
\citep{fm00, gebhardt.2000} which suggests
that black hole formation, galaxy evolution and AGN activity are all interlinked 
(\citealt{Schawinski.etal.2010}; \citealt{Somerville.etal.2008}; \citealt{Merloni.heinz.2008}). However, the 
$M-\sigma$ and $M-L$ relations show less scatter when only 
ellipticals and early type galaxies are included \citep{gultekin09}. Late type spirals introduce more 
scatter in the correlations. This may be due to a larger intrinsic scatter in the BH masses
of late type spirals or simply measurement uncertainities since late type spirals, in general, have smaller bulges.
They also show weaker AGN activity compared to the earlier type galaxies \citep{Ho.2008}.  Alternatively, In a study by 
\cite{beifiori.etal.2009} involving HST observations of 105 nearby galaxies spanning a wide range of Hubble types from
 ellipticals to late-type spirals, the estimated black hole masses upper limits appear to lie closer to the expected
 black hole masses in the most massive elliptical galaxies with values of $\sigma$ above 220 km/s than for galaxies with 
$\sigma$ in the range 90-220 \kms which appears to be consistent with a coevolution of supermassive black holes and 
galaxies driven by dry mergers.

It is not clear where LSB galaxies lie on the $M-\sigma$ plot. Their bulge velocity and disk rotation speeds suggest
that they lie below the $M-\sigma$ correlation for bright galaxies \citep{pizzella.etal.2005}. X-ray studies also suggest that
GLSB galaxies do not lie on the radio-x-ray correlation \citep{das.etal.xray.2009} and their black hole masses maybe quite
low \citep{naik.etal.2010}. In this paper we use detailed optical spectroscopy to determine 
the position of a sample of LSB galaxies on the $M-\sigma$ diagram. We observed the nuclear spectra of several
bulge dominated  LSB galaxies; for galaxies that showed AGN emission, we estimated both the black hole masses and also
the bulge velocity dispersion. In the following sections we present our observations, the results and discuss the 
implications of our findings.

\section{\sc\bf Sample Selection}

Our sample consists of nine large, bulge dominated LSB galaxies from \cite{schombert87}, of which eight have been observed 
further by \cite{schombert98}. The basic parameters of the galaxies are listed in Table \ref{tabparam}.  The LSB galaxies
in the Schombert sample were all \HI \ rich, giant, spiral galaxies that were derived from the UGC catalogue; they have 
systemic velocities that are less than $15,000$ \kms. We chose a subset of eight nearby galaxies from that sample that 
had $v_{sys}\le10,000$ \kms \ and appear to be bulge dominated galaxies with LSB disks. The last galaxy in our list, 
F568-6 (or Malin~2), has been observed by \cite{sprayberry95} and is also bulge dominated. 
The properties of individual galaxies are summarized below.\\
{\bf UGC~1378:~}This galaxy has a prominent bulge and a diffuse stellar disk. It is classified as an LSB galaxy with an
active nucleus by \cite{schombert98}. The bulge shows diffuse x-ray emission, possibly associated with an old stellar population
\citep{das.etal.xray.2009}.\\ 
{\bf UGC~1922:~}The disk of this galaxy is featureless but has a bright bulge that also shows diffuse x-ray emission. It is 
classified as an LSB galaxy with an active nucleus by \cite{schombert98}. UGC~1922 is also one of the few LSB galaxies that 
have a significant concentration of molecular gas \citep{oneil.schinnerer.2003}.\\
{\bf UGC~3968:~}Not much is known about this LSB galaxy except that it has a prominent bulge with a faint disk. The 2MASS image
reveals a bar associated with the bulge and two faint spiral arms. It is also classified as an LSB galaxy by \cite{schombert98} 
but it is not clear whether the galaxy has an AGN. \\
{\bf UGC~4219:~}Not much is known about this giant LSB galaxy either. According to \cite{schombert98}, the galaxy has a large bulge 
and an AGN. The LSB disk shows faint spiral arms. \\
{\bf UGC~6614:~}This is a well studied giant LSB galaxy \citep{deblok.etal.1995}. It is close to face on in morphology and has a 
large bulge surrounded by  a ring like feature (\citealt{Rahman.etal.2007}, \citealt{Hinz.etal.2007}). The disk has faint but tightly wound 
spiral arms \citep{pickering.etal.1997}. The bulge hosts an AGN that is bright at optical 
\citep{schombert98}, 
radio \citep{das.etal.2006} and x-ray \citep{naik.etal.2010} wavelengths. \\
{\bf UGC~6754:~}This galaxy has an LSB disk \citep{schombert87} and a prominent bulge but does not appear to have an AGN 
\citep{schombert98}. The disk has flocculent spiral arms and only patchy star formation \citep{amram.etal.1994}. \\
{\bf UGC~6968:~}Not much is known about this galaxy but it is described as an LSB galaxy having a prominent bulge and an AGN \citep{schombert98}. 
There are two faint spiral arms extending out into the disk. \\
{\bf UGC~7357:~}This galaxy has an LSB disk \citep{macarthur.etal.2003} and a bright bulge. But does not appear to have an AGN
\citep{schombert98}. The disk is fairly featureless. \\
{\bf F568-6 (Malin~2):~}This is also a relatively well studied giant LSB galaxy. It has an LSB disk (Schombert \& Bothun 1987),
prominent bulge and an AGN \citep{schombert98}. Like UGC~1922, it is one of the rare LSB galaxies that have a significant 
mass of molecular gas \citep{das.etal.2010}.

\section{\sc\bf Observations and Data reduction }

\subsection{HCT Data :}
The LSB galaxies were observed using the 2m Himalayan Chandra Telescope (HCT) at the Indian Astronomical 
Observatory (IAO), Hanle, which is remotely controlled from the Centre for Research and Education in Science and 
Technology (CREST), Indian Institute of Astrophysics (IIA), Bangalore. The spectra were 
obtained using a $11\arcmin\times1\farcs92$ slit (\#167l) in combination with a grism \#7 (blue region) and 
grism \#8 (red region) which cover the wavelength ranges of 3700--7200 \AA \ and 5500--9000 \AA \ with 
dispersions of 1.46 \AA \ pixel$^{-1}$ and 1.26 \AA \ pixel$^{-1}$ respectively. The spectral resolution is 
around $\sim 8.7$ \AA \ (398 \kms\ FWHM or $\sigma=169$ \kms\ at \ha) for grism \#7 and $\sim7$ \AA \ (
$\sigma = 136$ and 103 \kms\ at \ha\ and \ca2T\ respectively) for grism \#8. The slit was placed at the centre of the 
galaxy covering a central region of $\sim2''\times5''$ ($1''$ corresponds to 415 pc at a redshift of $\sim0.03$).

Data reduction was carried out using the standard tasks available within IRAF \footnote{Image 
Reduction \& Analysis Facility Software distributed by National Optical Astronomy Observatories, which 
are operated by the Association of Universities for Research in Astronomy, Inc., under co-operative 
agreement with the National Science Foundation} which includes bias subtraction, extraction of one 
dimensional spectra, wavelength calibration using the ferrous argon lamp for grism \#7 and ferrous neon 
lamp for grism \#8. The wavelength calibrated spectra were flux calibrated using one of the spectroscopic 
standards of \cite{oke90} observed on the same night and then corrected for the redshifts of the galaxies.
 Flux calibrated spectra were corrected for galactic extinction using \cite{schle98}. The spectra were not 
corrected for intrinsic dust extinction because LSB galaxies are known to have intrinsically less dust 
(\citealt{greeneho07}; \citealt{mei09}). The blue and the red spectra 
were combined together with the help of {\it scombine} within {\it specred} package using a 
suitable scale factor estimated at the flat continuum portions of the overlapping part of the spectra. A log of 
the observations are presented in Table \ref{tabobs} and the flux calibrated HCT spectra are plotted in 
Figure \ref{f:spec1}. We have used the HCT observations to identify the LSB galaxies that host AGN activity. The
black hole masses and bulge velocity dispersions were determined using SDSS data for all but one of these galaxies.

\subsection{SDSS DR7 Data :} Our pilot project on LSB galaxies was started in the year 2006 wherein careful selection of 
 objects for which SDSS data were unavailable was considered. But when the DR7 data was released to the astronomy community, 
we found  that a few galaxies from our sample were included. The resolution of the SDSS data is better
 ($\sim70$ \kms). Due to this, we have used SDSS DR7 data of our sample LSB 
galaxies for modeling/estimating the BH masses and velocity dispersions. For measuring the stellar velocity 
dispersion, SDSS offers a set of about 32 spectra of giant G and K stars of old open cluster M67. The stellar templates were
observed in an identical manner as the SDSS LSB spectra and hence effects arising due to template mismatch are minimal. 
SDSS spectra are observed through a fiber of $3''$ diameter which 
transforms into an area of 2.9 kpc diameter for a redshift of z$\sim0.021$. This area includes sufficient stellar light, nevertheless 
with stellar templates observed in the same setup, most of the stellar light would be removed after decomposition. In comparison, 
HCT spectra cover an area of $\sim5.4 \times 13.5$ kpc at the same redshift and the stellar templates were not observed in the same 
setup. Thus, for the estimation of BH masses after decomposing the broad and narrow components in the \ha \ region, and for the 
stellar velocity dispersion, we use SDSS spectra to give a conservative estimate of the above parameters.

\section{\sc\bf Spectral Decomposition}

The emission lines appear weak in our sample, but the \ha\ emission line is clearly present in several of the 
galaxies. In order to isolate the Balmer and major nebular emission lines better, we have decomposed the observed 
spectrum into different constituents. Major contribution to the observed spectra arises from the underlying 
stellar population, and we use high-resolution model spectra of Starburst99 \citep{leith99} after degrading 
it to the resolution of the observed spectra. While a composite spectrum based on stellar spectral libraries 
would have been more realistic, Starburst99 is simpler to use. 
The decomposition was executed only up to the wavelength of 7000 \AA \ as the Starburst99 high resolution model spectra are 
available only upto the above mentioned wavelength. 

The \ntwo \ $\lambda~6584$ and \othree \ $\lambda~5007$ lines along with \ha \ and \hb \ lines are used to 
calculate oxygen abundance assuming the empirical relation obtained by \cite{pp04}.
The oxygen abundances and hence metallicities are close to solar in value, particularly for F568-6 and UGC~6614 (see also 
\cite{mcgaugh94} for UGC~6614). The exception is the galaxy UGC~7357 which may have metallicity slightly less than solar. 
We have hence adopted solar abundances in applying the Starburst99 model. The method of \cite{mei09} was used for decomposition of 
stellar light, along with emission from the Fe\,{\sc i} and Fe\,{\sc ii} complexes, and in a few cases, a power law 
component. We use \cite{veron01} spectra of I Zw 1 to model the Fe\,{\sc i} and Fe\,{\sc ii} complexes.
Levenberg-Marquardt algorithm \citep{press93} was used for the decomposition.

Figure \ref{f:specfit1} shows the decomposed spectra plotted at the bottom of each plot. 
These spectra now show only the gas emission due to star formation and/or the active nucleus. 

According to \cite{schombert98}, in their sample of LSB galaxies, 95\% showed nuclear emission. Our sample 
consists of bulge dominated LSB galaxies and is a subset of the Schombert sample. Out of 
the 9 galaxies observed, we find that only 4 galaxies show broad \ha \ profiles emission, along with 
strong emissions from \ntwo, \stwo, \othree \ and \o, which hint the presence of an AGN. The 
line fluxes of various lines are presented in Table \ref{tabflux}.
We have adopted SDSS spectra for estimating the fluxes of broad and narrow components of \ha \ and hence in the
 calculation of BH masses, 
as well as stellar velocity dispersions. The SDSS spectra were taken through a fiber aperture of 3 arcsec in diameter (corresponding 
to 2.9 kpc at a redshift of 0.05). The broad and narrow components of \ha \ are seperated 
using the {\it fitprofs} task of {\sc IRAF}. Figure \ref{f:lfit1} shows the fits to the 
\ha \ line profiles. The broad \ha \ line in AGN spectra are generally asymmetric. The above procedure of multiple Gaussian
 fit results in slightly larger errors due to this. The \ha \ fluxes for broad and narrow components, \ha \ 
luminosity, full width at half maximum (FWHM) of the broad \ha \ lines are all listed in Table \ref{tabflux}. 

We have adopted the penalized-pixel fitting (pPxF) algorithm of \cite{mic.2004} for recovering the stellar velocity dispersion. 
pPxF is gauss-hermite parametrization (\citealt{marel.franx.1993}, \citealt{gerhard.1993}) that works in the pixel space. 
The reason is that in pixel space, it is easy to mask gas emission lines or bad pixels from the fit and the continuum 
matching can be directly taken into account \citep{mic.2004}. Also, the estimation of measurement errors are simplified. 
PPxF creates an algorithm, wherein, initial guesses for V (redshift, z) and $\sigma$ are provided. The model spectra are 
convolved with a broadening function using initial $\sigma$ values. $\chi^2$ is calculated for each dataset. The 
residuals or $\chi^2$ for each of the data points is perturbed and fed into a non-linear least squares optimization routine, 
in this case, monte-carlo optimization. The whole procedure is iterated to obtain V, $\sigma$ and gauss-hermite polynomials. 
The estimation of gauss-hermite polynomials becomes a problem if the observed velocity dispersion is of order ~2 
pixels or $\sigma$ $<$ 140 \kms \ (1 pix = 70 \kms \ for SDSS data). A detailed explanation of the procedure can be obtained from 
\cite{mic.2004}. 

\begin{figure}
 \begin{minipage}{150mm}
  \begin{center}
   \begin{tabular}{c}
    \includegraphics[width=9cm]{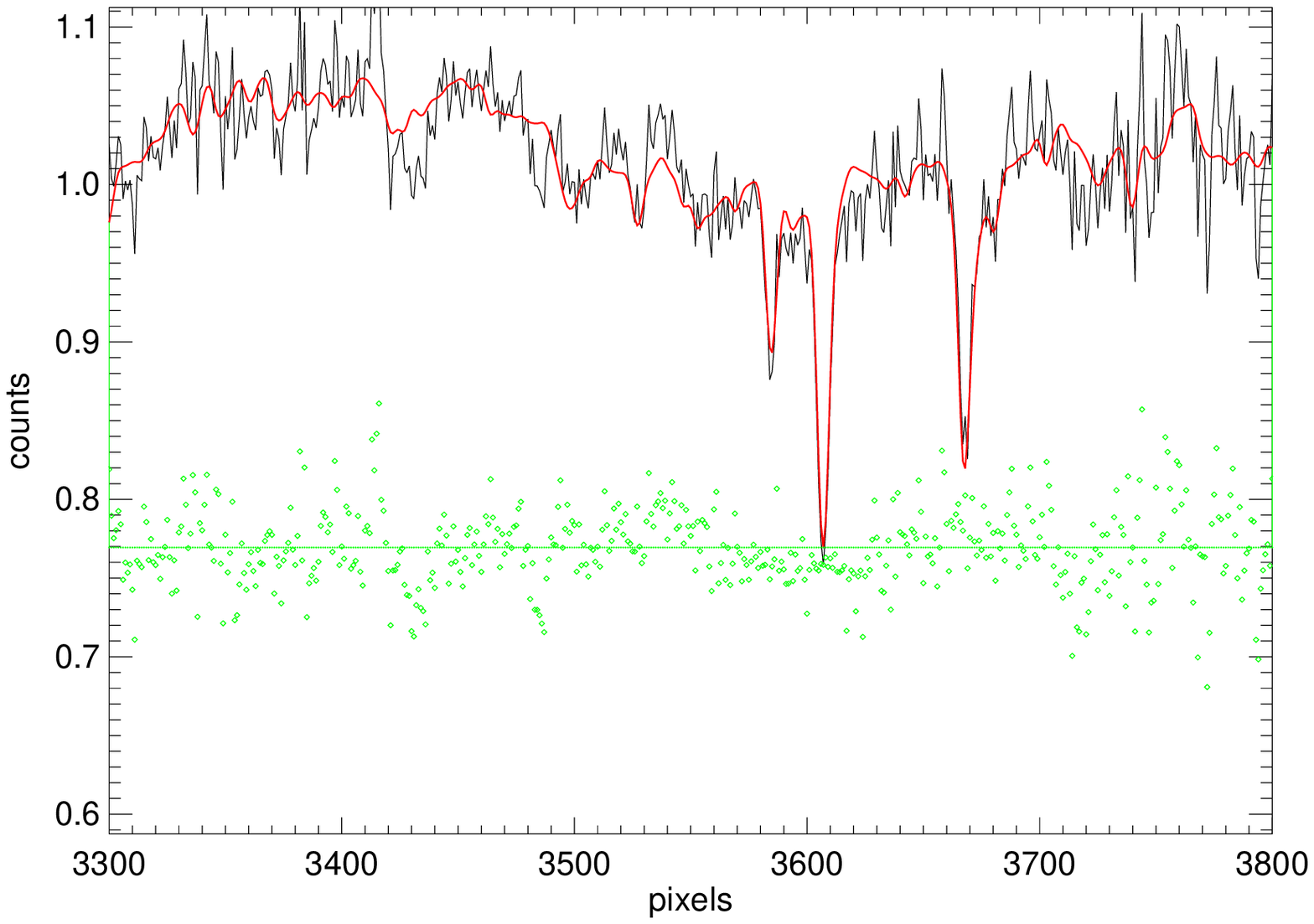} \\
    \includegraphics[width=9cm]{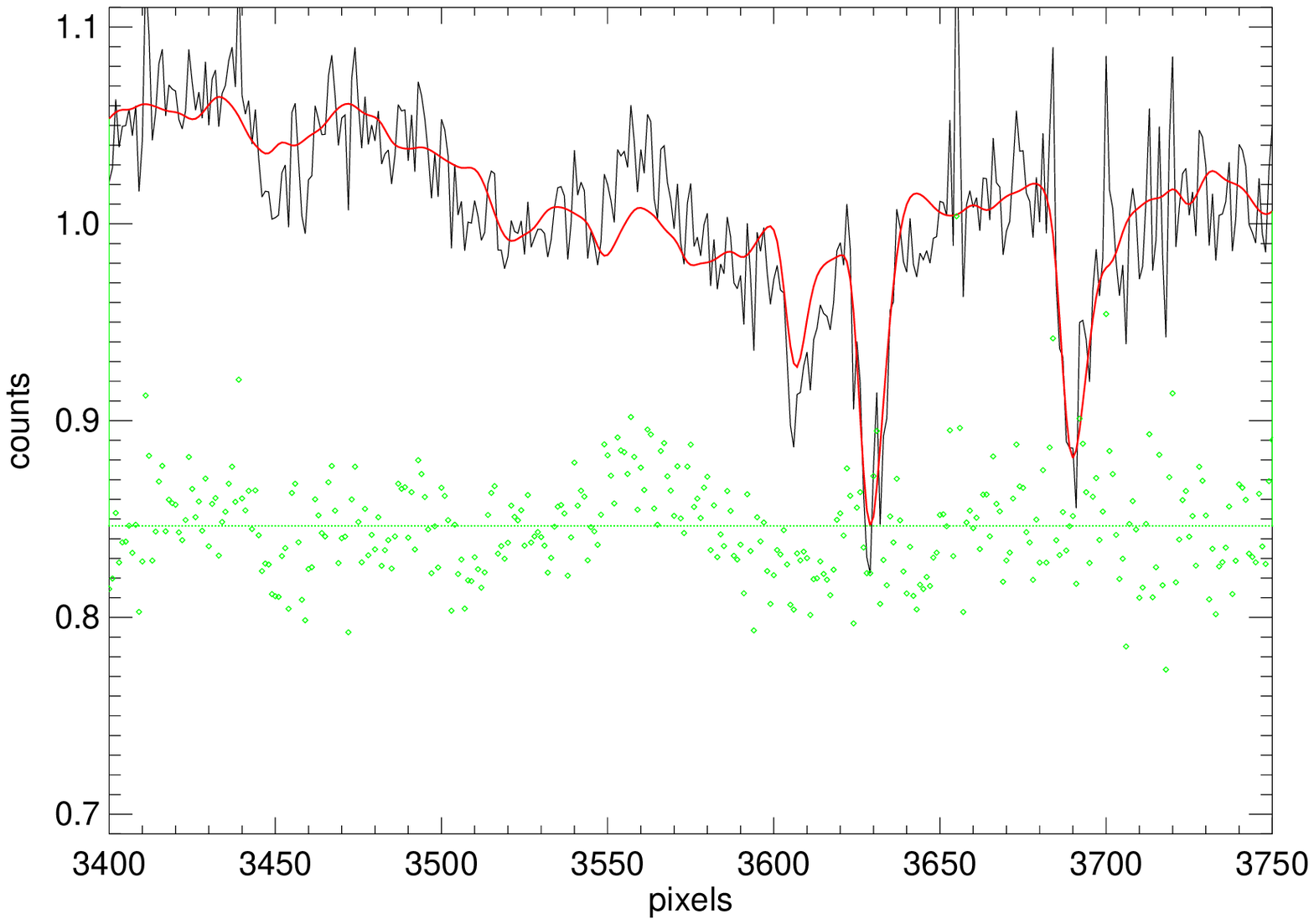} \\
    \includegraphics[width=9cm]{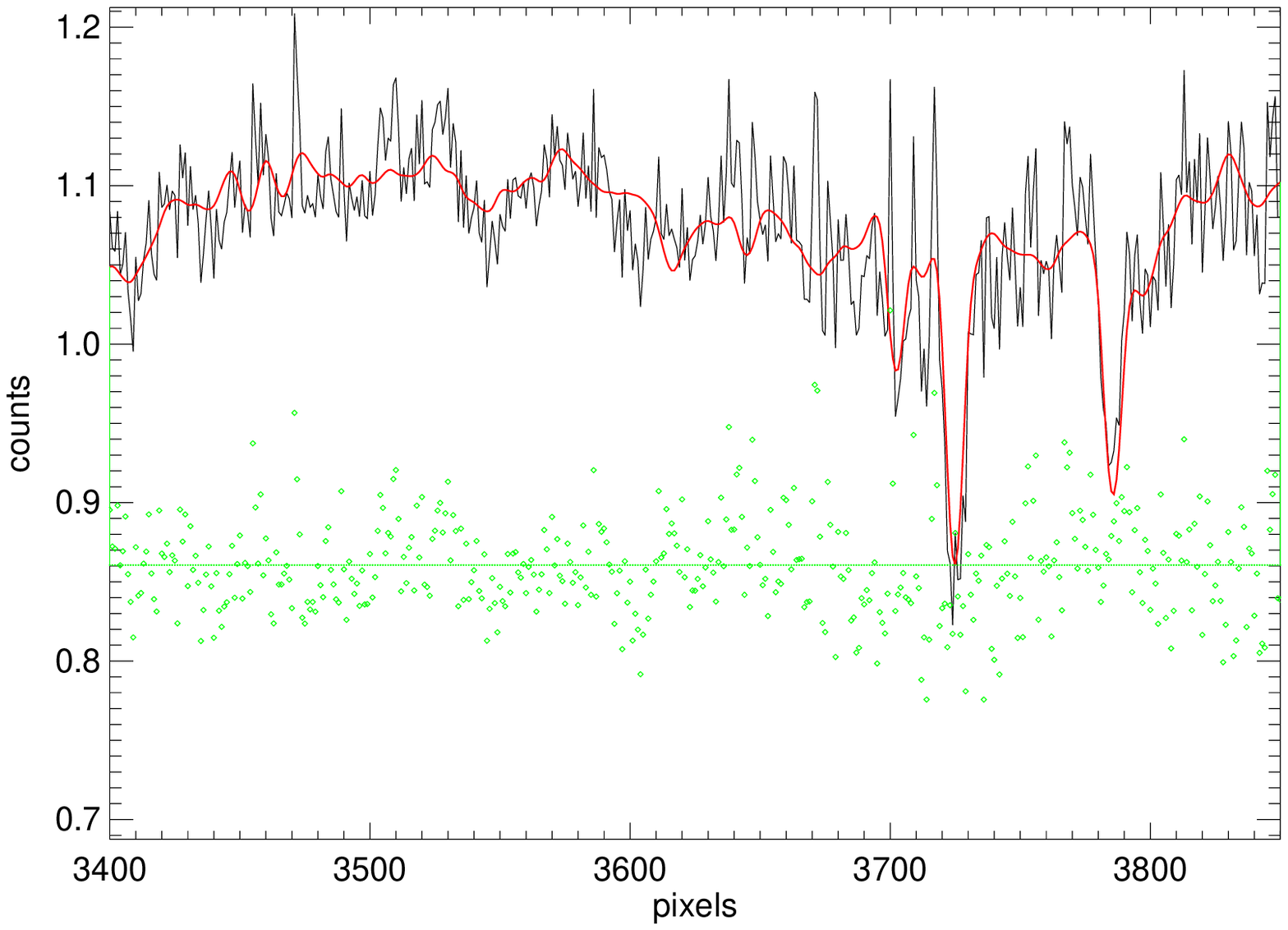} \\
   \end{tabular}
   \caption[]{Fits to the observed SDSS spectra using the pPxF code. The thick black line represents the observed spectra, 
the red line is the best fit SDSS template for the \ca2T region to estimate the velocity dispersion $\sigma_*$ for the 
three galaxies namely (from top to bottom) UGC 6614, UGC 6968 and F568-6. Measured $\sigma_*$ is 157, 196 and 209 \kms \
 respectively for the three galaxies. Green dots show the residues of the fit.}
  \label{f:ppxf}%
  \end{center}
 \end{minipage}
\end{figure}

With the availability of libraries with high spectral resolution stellar and galaxy spectra, templates 
can be carefully matched with the observed galaxy spectra. The 32 G and K giant stars of old open cluster M67 observed with 
the SDSS are used as stellar templates. Empirically, early K giants consistently provide the closest match to both the Mg{\sc i}b
and \ca2T regions of many AGN samples. Here, we have tried to measure the stellar velocity dispersion by fitting the \ca2T 
lines for these galaxies UGC 6614, UGC 6968 and F568-6. According to \cite{greeneho07}, the optimal spectral region for 
measuring $\sigma$ depends on the Eddington ratio, continuum level of the AGN, and redshift of interest which suggests that 
for z $<$ 0.05, \ca2T is the region of choice for Eddington ratio $\leq$ 0.5 for the most reliable measurement of
$\sigma$. The pPxF code was applied to only \ca2T region to estimate $\sigma$. Figure \ref{f:ppxf} 
displays the fits to the data; fitting error of $\sim10\%$ is estimated from the residuals.

\section{\sc\bf AGN Activity and Black Hole Masses}

\subsection{Emission Line Diagnostic Diagram}

\begin{figure}%
\begin{minipage}{150mm}

\includegraphics[width=15cm]{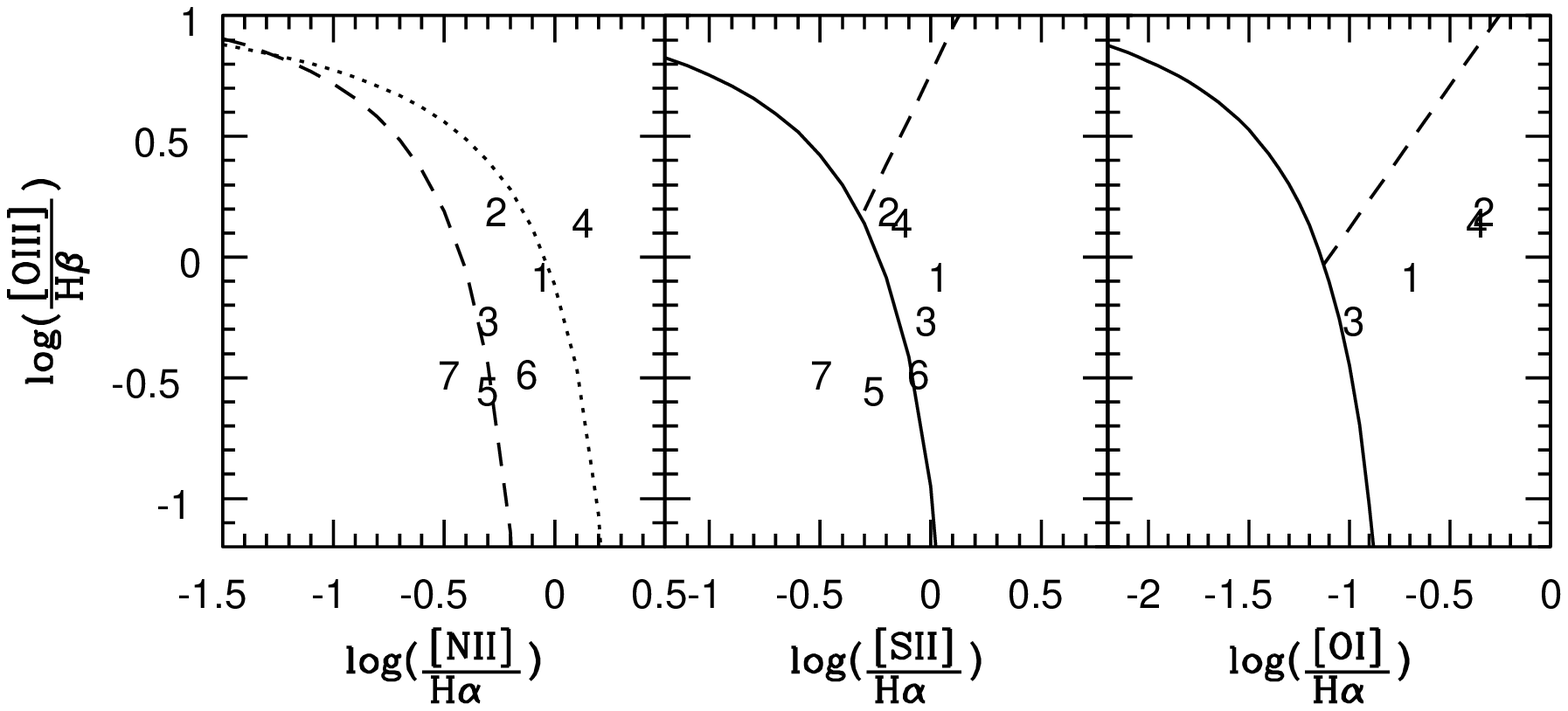} \\

1- UGC~1922; 2- UGC~6614; 3- UGC 6968; 4- LSBC~F568-6; 5- UGC~7357; 6- UGC~3968; 7- UGC~4219.\\

\caption[]{Diagnostic diagrams for the sample of LSB galaxies, observed from HCT. The dashed line in the leftmost figure, 
log({[O\,{\sc iii}]}/{\hb}) vs log({[N\,{\sc ii}]}/{\ha}), is the starburst-Seyfert demarcation
 taken from \cite{ka03}. The dotted line is the extreme starburst line of \cite{ke01}. The solid lines
showing the demarcation in log({[O\,{\sc iii}]}/{\hb}) vs log({[S\,{\sc ii}]}/{\ha}) and 
log({[O\,{\sc iii}]}/{\hb}) vs log({[O\,{\sc i}]}/{\ha}) is taken from \cite{ke01}. 
The straight line drawn in the centre and right plots is the Seyfert-LINER demarcation line taken from
 \cite{ke06}. The [O\,{\sc i}] emission was clearly detected only in four of the galaxies UGC 1922, UGC 6614, 
UGC 6968 and F568-6. These galaxies definitely host near-Seyfert kind of nuclei. UGC 7357 and UGC 4219 
clearly fall in the starburst regime, and the galaxy UGC 3968 might be hosting a starburst-AGN composite nucleus 
as seen from the diagnostic diagrams.}%
\label{f:dd}%
\end{minipage}
\end{figure}

It is interesting to investigate the positions occupied  by these LSB galaxies in the BPT diagnostic diagram 
(first given by \citealt{bpt81} and improved further by \citealt{vo87}, \citealt{ke01},
\citealt{ke06} and references therein). Figure \ref{f:dd} shows the diagnostic diagrams plotted for
\othree/\hb\ vs. \ntwo/\ha, \stwo/\ha\ and \o/\ha\ respectively. Also plotted in these diagrams are the demarcation lines
 between starbursts, Seyferts and LINERS obtained from \citealt{ke01}, \citealt{ka03}, \citealt{ke06}. From the plots, 
four galaxies appear strong candidates for AGN, showing LINER-like activity. 
Of these, the galaxies F568-6 and UGC~6614 also show a high value of \othree/\hb $\sim$1.5 when compared to other galaxies 
in the sample and could be Seyfert-like. It may be noted that the \othree/\hb\ ratio for low-mass AGNs selected from SDSS 
is 1.9 for one kind of sample \citep{greeneho07}. The width of the broad \ha \ lines observed in these galaxies is below
$\sim2000$~\kms \ indicating that they both belong to the class of narrow-line Seyfert 1s (NLS1s) galaxies.

\subsection{Stellar Velocity Dispersion $\sigma_{*}$}

The stellar velocity dispersion $\sigma_*$ is measured using the \ca2T \ lines adopting the pPxF code of \cite{mic.2004}. 
We could detect the \ca2T \ lines at 8542 \AA \ and 8662 \AA \ in some of our galaxies such as UGC 6614 and
F568-6. Hoever, due to template mismatch, we could not estimate the $\sigma_*$ using the HCT spectra. Hence we used the 
SDSS spectra for measuring $\sigma_*$. The spectral resolution of SDSS spectra are about 4 \AA, which amounts to a 
velocity width of about 140 \kms\ and $\sigma=70$ \kms.
The derived values for UGC 6614, UGC 6968 and LSBC F568-6, are in the range 150--210 \kms. The SDSS spectrum is not 
available for the galaxy UGC 1922 and hence we could not estimate the stellar velocity dispersion for this galaxy. 
We estimate $\sigma_* \sim157$, $\sim196$ and $\sim209$ km\,s$^{-1}$ for UGC 6614, UGCC 6968 and f568-6, 
respectively using SDSS spectra. These values are shown in Table \ref{tabmbh}. Figure
\ref{f:ppxf} shows the fits to the observed data obtained using the pPxF code.
 
There is some discussion in the literature on the need to reduce the observed stellar velocity dispersions to a 
uniform system, since the observed values are averaged over the slit or aperture size which will translate into
 different sizes on the face of the galaxy with respect to the bulge scale length. 
\cite{jfk95} transform the values to the equivalent of an aperture of radius $r_e$/8, 
where $r_e$ is the effective bulge radius. The effective bulge radius is about 4\farcs2 for UGC 6968 as calculated 
by \cite{gavazzi00} using near-IR $H$ band image of the galaxy. 
Though \cite{mcgaugh94} have attempted only to 
fit the disk, the bulge is visible in their surface brightness profile plots for the galaxies UGC~6614 and F568-6. 
\cite{deblok95} have also obtained the surface brightness profile for UGC~6614 and fit only the disk. 
The effective bulge radius r$_e$ for the galaxies UGC~1922, UGC~6614 and F568-6 are not available in the literature. 
On the other hand, as pointed out by \cite{fm00}, the applied corrections for the velocity dispersion $\sigma_*$ 
are very small, the maximum correction being $<5\%$. \cite{greeneho05} and \cite{piz04} find that radial dependence of 
$\sigma_*$ is flat with less than $7\%$ correction for early-type galaxies. 
Following these arguements, we have not applied any correction based on $r_e$ to $\sigma_*$, the spectra are 
extracted from a region of $2''\times5''$. 

\subsection{The Mass of Central Black Hole ($M_{BH}$)}

The blackhole masses are calculated using the equation given in \cite{greeneho07} using \ha \ luminosity and FWHM. 
The masses  are $\sim0.3~\times10^6\ M_\odot$ for the galaxies UGC 1922, UGC 6968 and 
F568-6, and lie on the lower mass tail of the low-mass blackhole sample of \cite{greeneho07} which has a median 
mass of $1.3~\times10^6$ M$_\odot$. UGC 6614 has a slightly higher BH mass of $3.8~\times10^6\ M_\odot$. 
The average $L_{\rm bol}/L_{\rm Edd}$ ratio calculated for
their sample is about 0.4 and suggests that their sources are radiating at high fraction 
of Eddington limit \citep{greeneho07}. We find the values of
0.18, 0.023, 0.046 and 0.106 for UGC 1922, UGC 6614, UGC 6968 and F568-6 respectively. 
While these are lower compared to the median for the sample of \cite{greeneho07}, they are within their observed range. 

We could identify clear signature for AGN in 4 out of 9 objects in our sample which agrees with the high (50\%) 
occurrence of AGN found by \cite{schombert98} in LBGs. 
\\

\subsection{Interesting case of UGC 6614} 
\begin{figure}
\begin{minipage}{150mm}
\centering
\includegraphics[width=10cm]{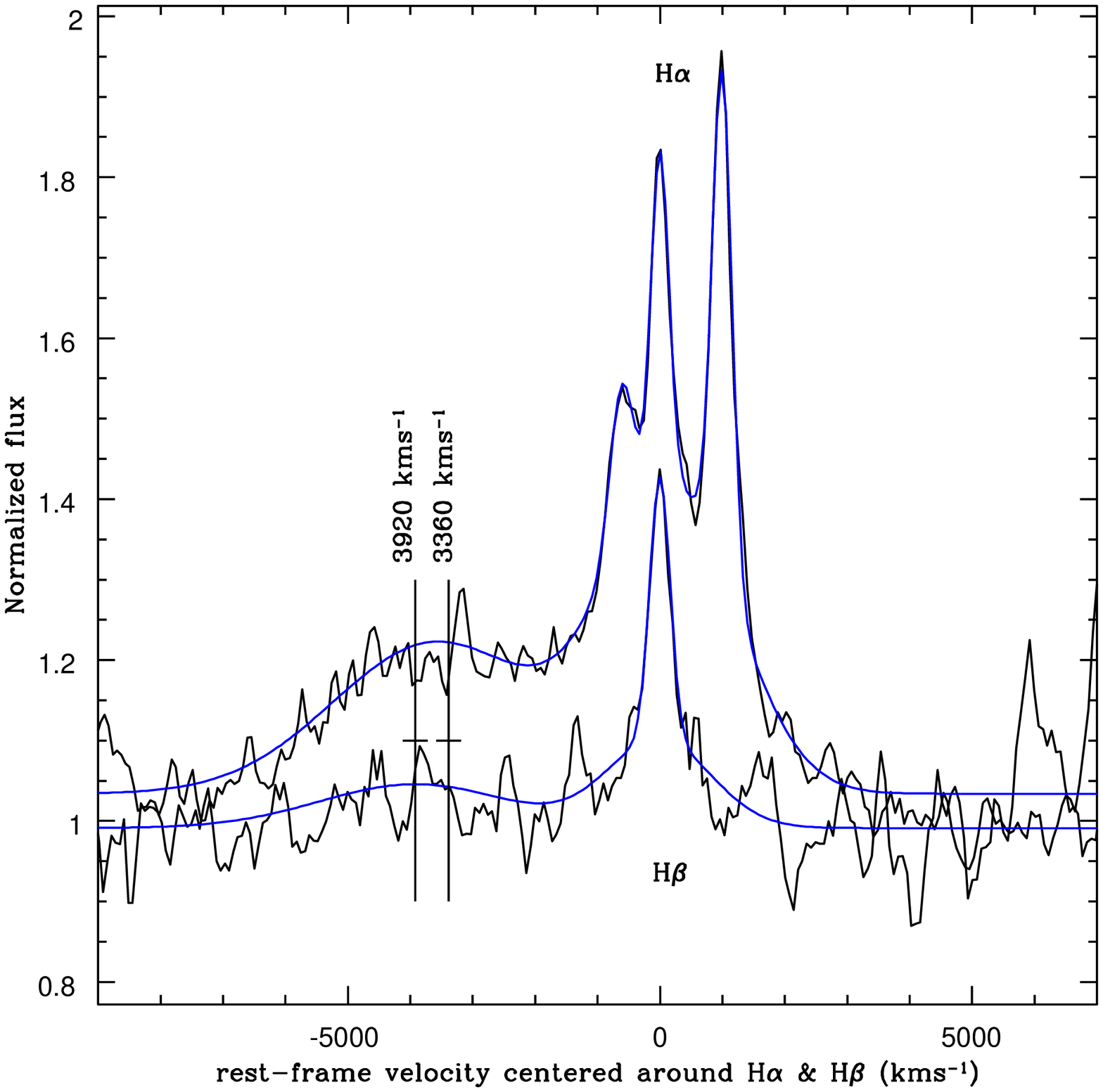} 
\caption[]{Line fits for the region around \ha \ and \hb \ for the galaxy UGC 6614. The x-axis is interms of velocities. 
A blue bump is clearly noticeable around \ha \ and \hb \ which is centered at $-3600\pm300$ \kms \ and moving at a velocity 
of $\sim3600$ \kms \ towards us.}
\label{f:ha_hb_outflow}
\end{minipage}
\end{figure}

The emission line spectra of UGC 6614 obtained after the decomposition of stellar light shows an interesting feature. A
bump is noticed at the blueward of the \ha \ emission from the galaxy. This blue bump is noticeable in the observed flux 
calibrated spectra before spectral decomposition but is clearly seen after the decomposition. The blue bump is the excess
 emission at \ha \ which could be arising from ionised gas travelling at speeds $\sim3600$ \kms \ towards us, centered at 
3920 \kms. The blue bump is also noticed blueward of \hb \ and the velocities with which the gas streaming out towards us 
$\sim3600$ \kms \ (median value of \ha \ and \hb) centered at 3360 \kms \ from \hb, similar to \ha.
This emission at \ha \ and \hb \ wavelengths overlap indicating the feature to be real as shown in Figure 
\ref{f:ha_hb_outflow}.
\cite{das.etal.radio.2009} detected a compact core and a one-sided radio jet in UGC 6614 from 610 MHz map. An extended 
feature is also
 indicated in a low resolution VLA map at 1420 MHz \citep{das.etal.radio.2009}. The blue shifted ionized gas emission 
could indicate a jet or hotspot along the line-of-sight. Similar asymmetric blue bump of the [OIII] lines were detected 
from a bunch of Type 1 - Type 2 Seyferts from SDSS DR2 sample by \cite{greeneho05}. These wings indicate radial motions in 
the NLR, associated with an outflow. The outflowing components are principally responsible for imparting supervirial 
motions to the gas and originate from a more compact region closer to the centre \citep{greeneho05}.

\subsection{The $M_{BH}-\sigma_*$ \ Plot}
\begin{figure}%
\begin{minipage}{150mm}
\centering
\includegraphics[width=10cm]{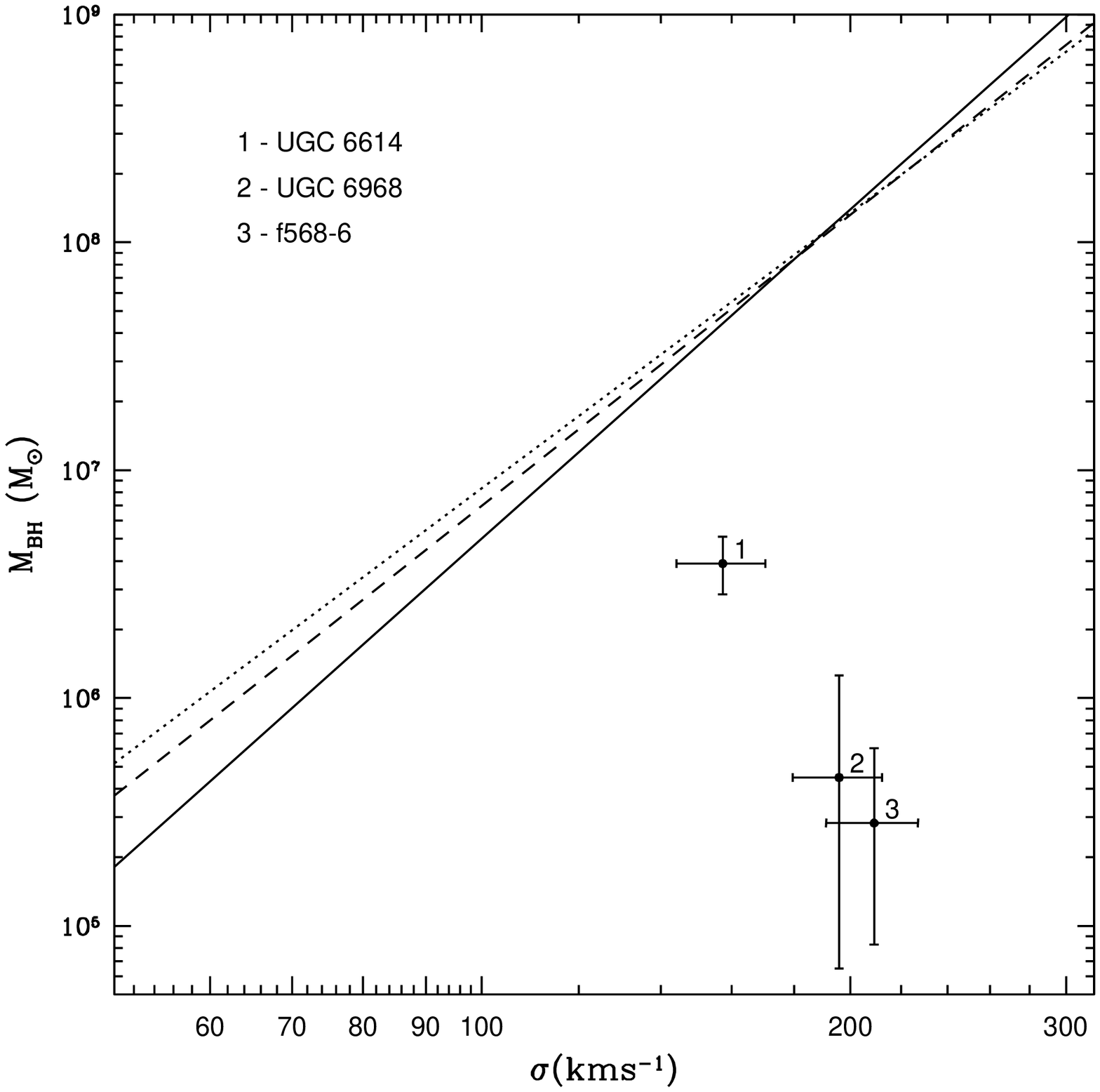} 
\caption[]{$M-\sigma$ plot for the LSBs. Points represent the $M_{BH}$ and velocity dispersion of the 3 LSBs
  UGC 6968, UGC 6614 and F568-6. The solid line is the $M_{BH}$-$\sigma_*$ relation taken from 
\cite{fm00}, the dashed line is from \cite{Tremaine.etal.2002} and dotted line is taken from \cite{gultekin09}. The $M_{BH}$ and 
$\sigma_*$ are estimated from the SDSS spectra.}%
\label{f:msig}%
\end{minipage}
\end{figure}

The \ca2T line widths and masses of blackholes for the three galaxies in our sample are shown in the  
$M_{BH}-\sigma_*$ plot in Figure \ref{f:msig}. Also plotted in the figure are the linear regression lines given by
 \cite{gultekin09},
 \cite{Tremaine.etal.2002} and \cite{ff2005} (dotted, dashed and solid lines respectively) for $M_{BH}$ against $\sigma_*$. 
The low-mass AGNs hosted within LSB galaxies occupy the region just below the lowest mass blackhole of Circinus 
galaxy from the sample of 
\cite{gultekin09}, well below extrapolations of high-mass blackholes. On the other hand, three AGNs in LSBs 
observed by \cite{mei09} in the blackhole mass range of $2.8-20\times 10^6\ M_\odot$ lie closer to the
 \cite{Tremaine.etal.2002} relation, though systematically lower. It would be of interest to study more LSBGs and low 
luminosity AGN of \cite{greeneho07} for a better understanding of faint luminosity end of $M_{BH}-\sigma_*$ 
relation.

\section{Discussion}

 {\bf 1.~Intermediate Mass Black Hole (IMBH) in GLSB galaxies:} One of the main results of our spectroscopic study 
is the detection of broad \ha \ emission in GLSBs and the subsequent estimation of nuclear black hole masses from SDSS 
spectra in bulge-dominated GLSB galaxies. The AGNs fall in Seyfert-LINER region in the diagnostic diagram (refer 
figure \ref{f:dd}). 

We obtained masses $\sim3~\times10^{5}~M_{\odot}$ for three GLSB galaxies in our sample, which fall in the IMBH 
range rather than the SMBH range. 
A higher blackhole mass is estimated for UGC 6614 which is $\sim3.8~\times10^{6}~M_{\odot}$ and a similar estimate was 
given earlier by \cite{das.etal.xray.2009} based on a low resolution optical spectrum from \cite{sprayberry95}. 
Another estimate of $\sim1.2\times10^{5}~M_{\odot}$ was derived later from AGN  X-ray variability studies by 
\cite{naik.etal.2010}, which is lower than the present estimate. 
It must be borne in mind that these estimates are fairly approximate as they are based on the assumption 
that the gas in the broad line region in the AGN is in virial equilibrium \citep{Kaspi.etal.2000}.

IMBHs are fairly rare in the galactic nuclei and have been detected mainly in late type spirals 
(\citealt{Filippenko.Ho.2003}; \citealt{Greene.Ho.2004}; \citealt{Satyapal.etal.2007}), nearby galaxies 
\citep{seth.etal.2010} or dwarf galaxies \citep{Barth.etal.2004}. They are difficult to detect dynamically at large 
distances; hence AGN activity is one of the main methods through which we detect them. The presence of IMBHs in GLSB
galaxies is surprising as their bulges are well developed; in fact a SMBH would be far more typical for
these bulges. This suggests that the lack of disk evolution in these extreme late type galaxies has affected the
evolution of their nuclear BHs. Galactic disk activity contributes to the growth of SMBHs through gas inflow, 
star formation and mass accumulation in the nuclei of spiral galaxies as observed in bulge dominated, 
star forming early type spirals. Large scale disk instabilities such as bars and spiral arms 
exert gravitational torques that funnel gas into galaxy centers leading to nuclear star formation and the build-up
of central mass concentrations (e.g. \citealt{friedli.benz.1993}). This can result in the growth of nuclear black holes
and bulges in galaxies \citep{kormendy.kennicutt.2004}. This process of disk evolution leading to the growth of central 
mass concentrations in galaxies is prevented from happening or slowed down when there is a dominant dark matter halo
\citep{ostriker.peebles.1973}. Thus, the lack of disk evolution and relatively low mass of the black hole may share 
the same origin - which is the presence of a dominant dark halo in the galaxy. 
 
{\bf 2.~Constraining the $M-\sigma$ relation for extreme late type galaxies:} In the past ten years the $M-\sigma$ 
and $M-L$ relations have become established benchmarks for galaxy and BH evolution theories. Our present work affects
these correlations in two ways; first it helps constrain the low mass end of the $M-\sigma$ correlation and secondly it
helps to constrain the scatter in the plot \citep{Gultekin.etal.2010}. The low mass end of the 
$M-\sigma$ relation is populated by late type galaxies or dwarf galaxies; many are often outliers in the plot. 
Our present work shows that extreme late type galaxies are also fairly offset from the main 
$M-\sigma$ line (Figure \ref{f:msig}). It also suggests that dwarf galaxies and extreme late type galaxies have different 
evolutionary paths compared to early type galaxies and the more massive ellipticals at the high SMBH end 
of the $M-\sigma$ relation. Models of galaxy evolution thus need to incorporate late type systems such as GLSB
galaxies in their overall picture. 

The late type spirals also increase the scatter in the correlation which is tighter when only ellipticals and early 
type spirals are included (\citealt{Tremaine.etal.2002}, \citealt{beifiori.etal.2009}). Many theoretical studies have been
 undertaken to explain the $M-\sigma$ correlation and predict the 
high mass end of the plot \citep{natarajan.treister.2009}. \cite{dallabonta.etal.2009} carried out HST observations of 
three brightest cluster galaxies (BCGs) and estimated masses of SMBHs to be $\sim10^9$ $M_{\odot}$ present in these BCGs. 
While for one galaxy, SMBH mass correlates well in the $M-\sigma$ and $M-L$ plot at the high mass end, the other two 
galaxies show inconsistencies with the two relations. The sample is small to derive any conclusions, but hints that there 
could be scatter in the SMBH scaling relations at the high mass end as well \citep{dallabonta.etal.2009}.   

However the low mass end does not appear to have a clear 
cutoff according to most models \citep{volonteri.natarajan.2009}. In fact, we could be missing observationally 
a large fraction of the lower mass BHs in the centres of galaxies. Thus there is an increase in the scatter at the low 
mass end of the correlation. This scatter could be due to measurement errors or could be intrinsic to the BH evolution 
processes in the galaxies, themselves \citep{Volonteri.2007}. Our present 
study is thus important for understanding the overall trends in the low mass end of the BH mass spectrum.

{\bf 3.~AGN evolution in late type galaxies:} Studies have shown that the space density of high luminosity AGNs
peak at redshift of $z\sim 2$; this is also the redshift at which the most massive SMBHs were formed in galaxies 
(\citealt{Cowie.etal.2003}; \citealt{Hasinger.etal.2005}). However, in the local universe the 
most rapidly growing BHs appear to be those in the lower mass range of $10^{6}-10^{7}~M_{\odot}$ 
\citep{Goulding.2010}. Also, studies of nearby galaxies show that it is the most massive BHs in late type galaxies
that are growing at the present epoch \citep{Schawinski.etal.2010}. The GLSB galaxies in our sample
fall into the latter category as they have large bulges; though they have lower BH masses, they appear to be accreting 
and hence luminous in the optical domain.

 {\bf 4.~Decoupled Bulge-Disk Evolution in GLSB galaxies:} As suggested by \cite{das.etal.xray.2009}, the bulges of 
GLSB galaxies appear to be very evolved compared to their disks. In general bulges form in two ways; one is through 
repeated galaxy mergers or accretion events that lead to the formation of a central spheroidal mass distribution
\citep{springel.etal.2005}. The 
second is through secular evolutionary processes where disk instabilities lead to bars, spiral arms, gas infall and the evolution 
of a disky pseudobulge \citep{kormendy.kennicutt.2004}. These processes result in disk star formation and enhanced disk structure,
both of which are not observed in most GLSB galaxies; instead their disks are metal poor and often fairly featureless. So the 
bulges in GLSB galaxies probably formed in a different way; one possibility is that galaxy mergers resulted in spheroidal bulges 
and then the disks were rebuilt from accreted gas \citep{springel.hernquist.2005}. Such an evolutionary scenerio would lead 
to a bulge that is relatively decoupled from its disk or its central black hole.

{\bf 5.~BHs in Halo Dominated Galaxies:} Although the correlation of BH mass and galaxy properties is now
well established, it is still not clear exactly what regulates black hole growth (e.g. \cite{booth.schaye.2010} and
references therein). Mass accretion close to the black hole, bulge mass and the mass of the dark matter halo are
some of the factors important for regulating black hole growth in galaxies. It is not clear which factor is the most 
important or whether all the processes play a role. Several theoretical studies have explored how the potential 
of the dark halo may regulate bulge evolution and black hole growth (\citealt{booth.schaye.2010}; 
\citealt{xu.etal.2007}; \citealt{silk.rees.1998})
and there are observational studies that indicate a correlation between the dark halo and black hole mass 
(\citealt{baes.etal.2003}; \citealt{ferrarese.2002}; \citealt{pizzella.etal.2005}). On the other hand, some studies 
show that nuclear black holes masses do not correlate with the dark halo matter haloes of galaxies and dark matter 
gravity is not directly responsible for black hole growth (\citealt{kormendy.bender.2011}; \citealt{ho.2007}).
Such studies suggest that SMBHs co-evolve with classical bulges or ellipticals only. GLSB galaxies are 
halo dominated and often bulge dominated as well. Hence they are ideal systems to study the dark halo-BH
relation and this should be investigated in future studies.

\section{\sc\bf Conclusion}

\begin{enumerate}
 \item
The paper presents spectroscopic observations of the nuclear regions of nine low surface brightness 
galaxies observed in the wavelength range 3700--9000 \AA. The stellar light has been subtracted from the nuclear 
spectra to obtain only gas emission spectra.  
Broad \ha \ lines along with strong [N{\sc ii}], [S{\sc ii}], [O{\sc i}] 
lines are detected in four galaxies namely UGC 1922, UGC 6614, UGC 6968 and F568-6 confirming the presence of AGN
activity in these LSB galaxies. SDSS spectra of three galaxies namely, UGC 6614, UGC 6968 and F568-6 are used 
to estimate the BH masses and stellar velocity dispersion as the resolution of the SDSS spectra is good enough ($\sim70$ \kms) 
to give a conservative estimate on the above parameters.

\item
The BPT AGN diagnostic diagram was
created using the emission line ratios. It is clearly seen that the above four galaxies lie in the AGN
regime and more closely, in the Seyfert regime. UGC 3968 might also host a starburst-AGN composite nucleus. 

\item
The broad \ha \ line widths (900--2500 km\,s$^{-1}$) and luminosities ($10^{39}$ erg\,s$^{-1}$) are
used to deduce the nuclear blackhole masses in the aforementioned galaxies; the masses for three galaxies are 
$\sim3~\times10^{5}~M_{\odot}$ and for UGC 6614, the BH mass is estimated to be about $3.8\times10^{6}~M_{\odot}$. 
The masses suggest that the nuclei of LSB galaxies have IMBHs rather than the SMBHs
found in the centres of brighter galaxies. UGC 6614 also shows an interesting feature of a blueshifted bump of \ha \ emission 
which can be attributed to outflow of gas travelling at speeds of 3600 \kms \ towards us. The blue bump feature is  
detected in the \hb \ region as well.
There could be more such AGN in the sample that may be identified through other means 
such as the reverberation technique with improved sensitivity.

\item
The stellar velocity dispersion, $\sigma_*$ is measured 
for the three galaxies UGC 6614, UGC 6968 and F568-6; the values lie between 150--210 km\,s$^{-1}$. The three low-mass BHs lie 
below the standard line in the $M-\sigma_*$ plot and lower than the ones studied by \cite{mei09}. 
\end{enumerate}

\section{\sc\bf Acknowledgements}
We would like to thank the anonymous referee for suggestions and clarifications which enhanced the quality of the 
paper immensely. We thank Dr. J.E. Greene for providing the SDSS templates and IDL 
fitting routine, Dr. Monica Valluri for suggestions, fruitful discussions 
and for introduction to pPxF algorithm, and Dr. T. Sivarani for help and 
discussions for the pPxF fitting routines. \\
We thank the staff of  IAO and CREST for their help during the observations. \\
This research has made use of the NASA/IPAC Extragalactic Database (NED) which is operated by the
 Jet Propulsion Laboratory, California Institute of Technology, under contract with the National Aeronautics and 
Space Administration.\\
The SDSS is managed by the Astrophysical Research Consortium (ARC) for the Participating 
Institutions. The Participating Institutions are The University of Chicago,Fermilab, the Institute for Advanced 
Study, the Japan Participation Group, The Johns Hopkins University, Los Alamos National Laboratory,the 
Max-Planck-Institute for Astronomy (MPIA), the Max-Planck-Institute for Astrophysics (MPA), New Mexico State 
University, Princeton University, the United States Naval Observatory, and the University of Washington. \\
RS would like to thank the University Grants Commission (UGC) for their UGC-CSIR NET fellowship given by the 
Government of India.

\bibliographystyle{apj}
\bibliography{ram}

\begin{figure}%
\begin{minipage}{150mm}
\centering
\includegraphics[width=11cm]{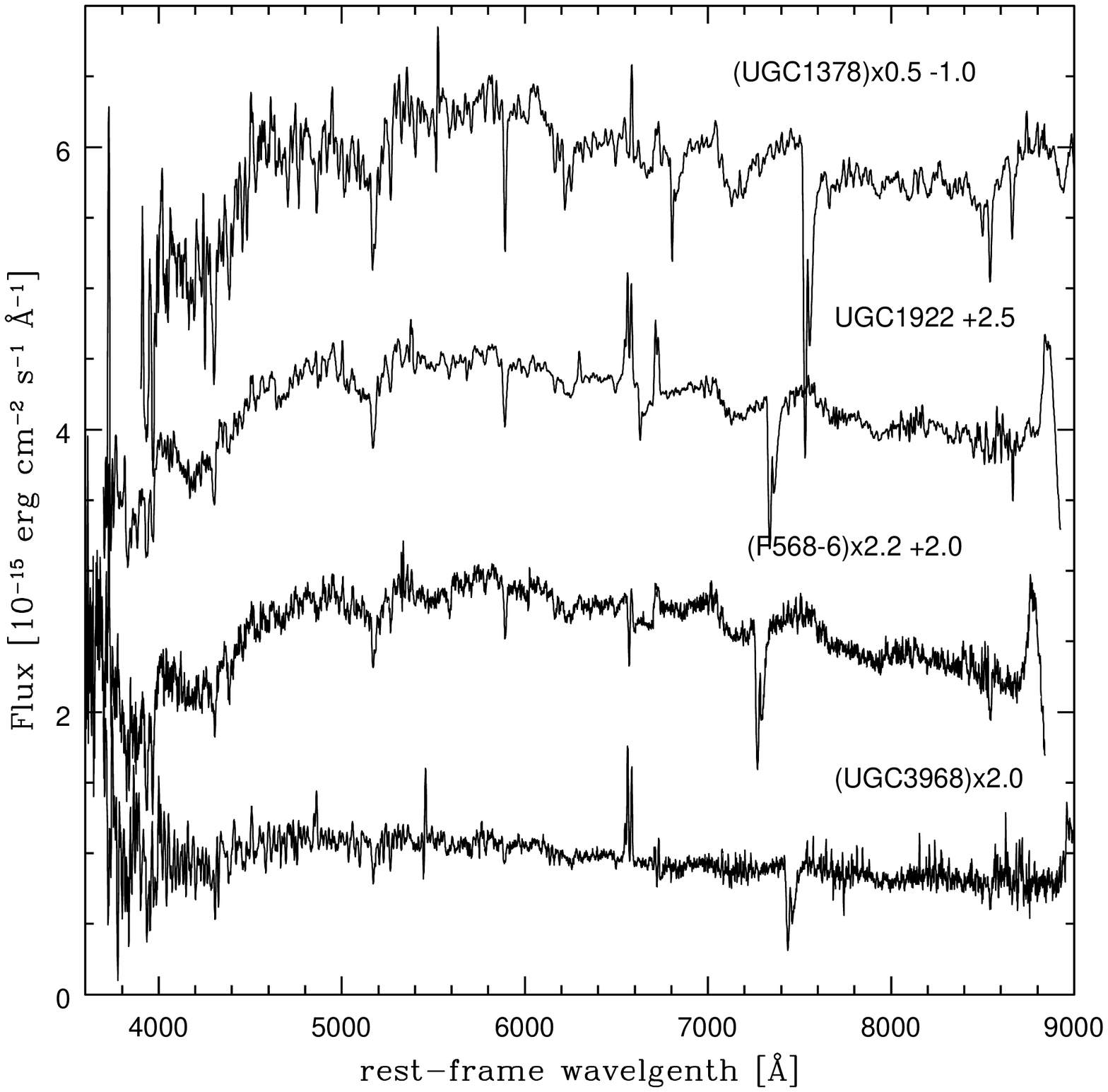}%
\caption[]{Spectra of the LSBs observed from HCT using the blue (\#7) and red (\#8) grism and combined together 
using the {\it scombine} task of IRAF. The x-axis represents the rest-frame wavelength. The spectra are arbitrarily 
displaced for clear viewing. }%
\label{f:spec1}%
\end{minipage}
\end{figure}

\begin{figure}%
\ContinuedFloat
\begin{minipage}{150mm}
\centering
\includegraphics[width=11cm]{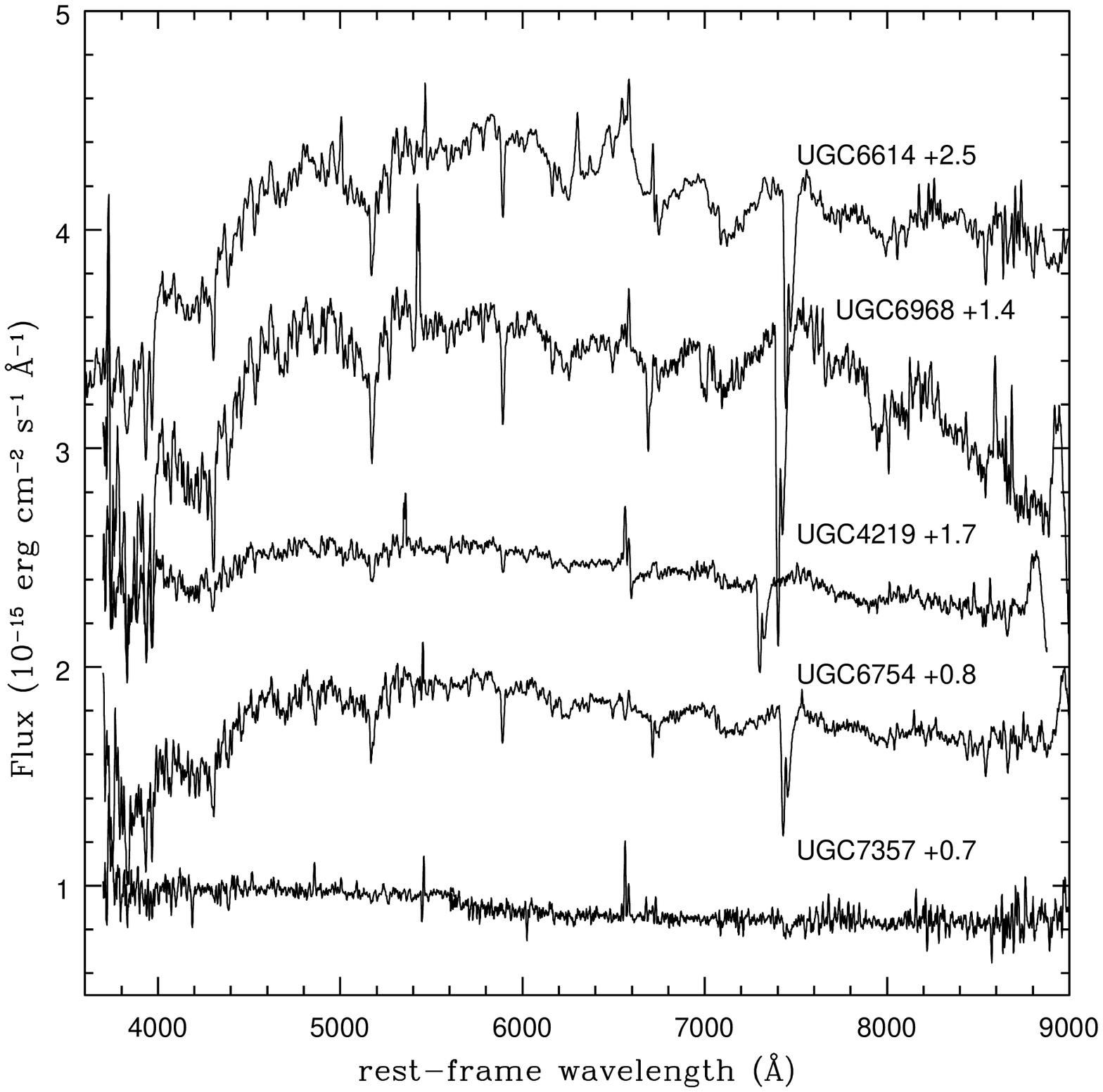}%
\caption[]{ - \textit{Continued}}%
\label{f:spec2}%
\end{minipage}
\end{figure}

\begin{figure}%
\begin{minipage}{150mm}
\centering
\begin{tabular}{cc}
UGC 1378 & UGC 1922 \\
\includegraphics[width=7cm]{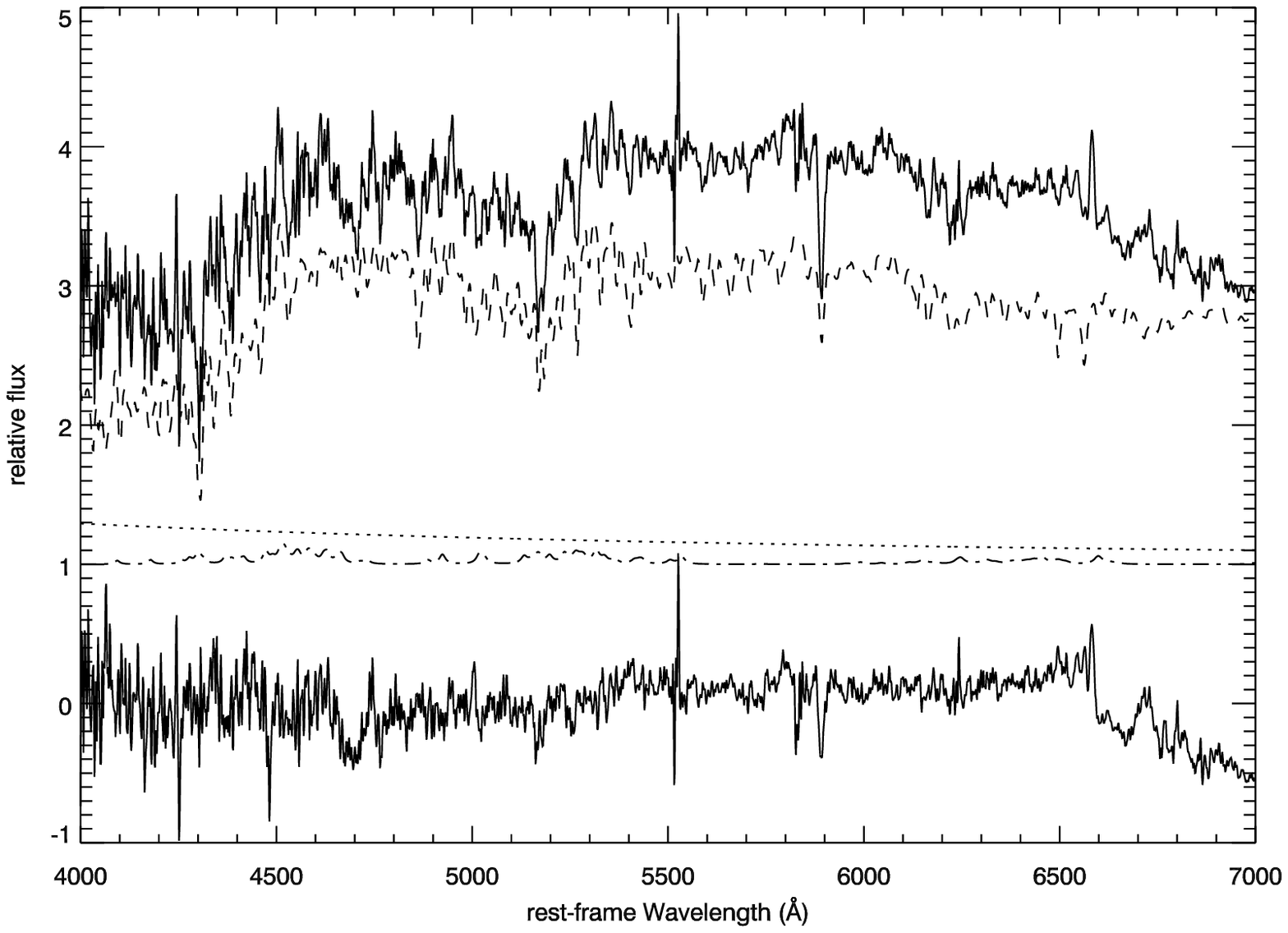}
&
\includegraphics[width=7cm]{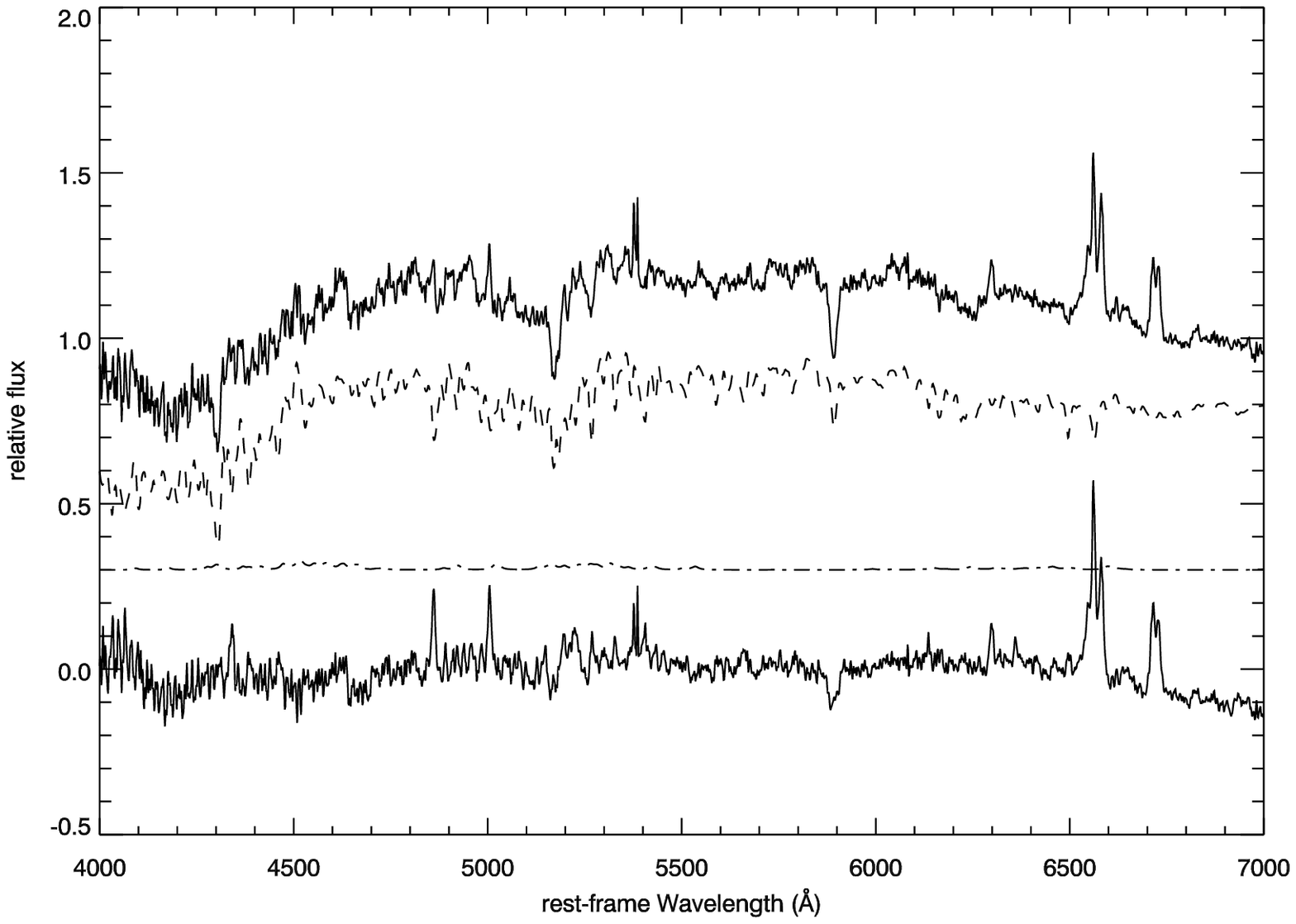} \\
UGC 3968 & UGC 4219 \\
\includegraphics[width=7cm]{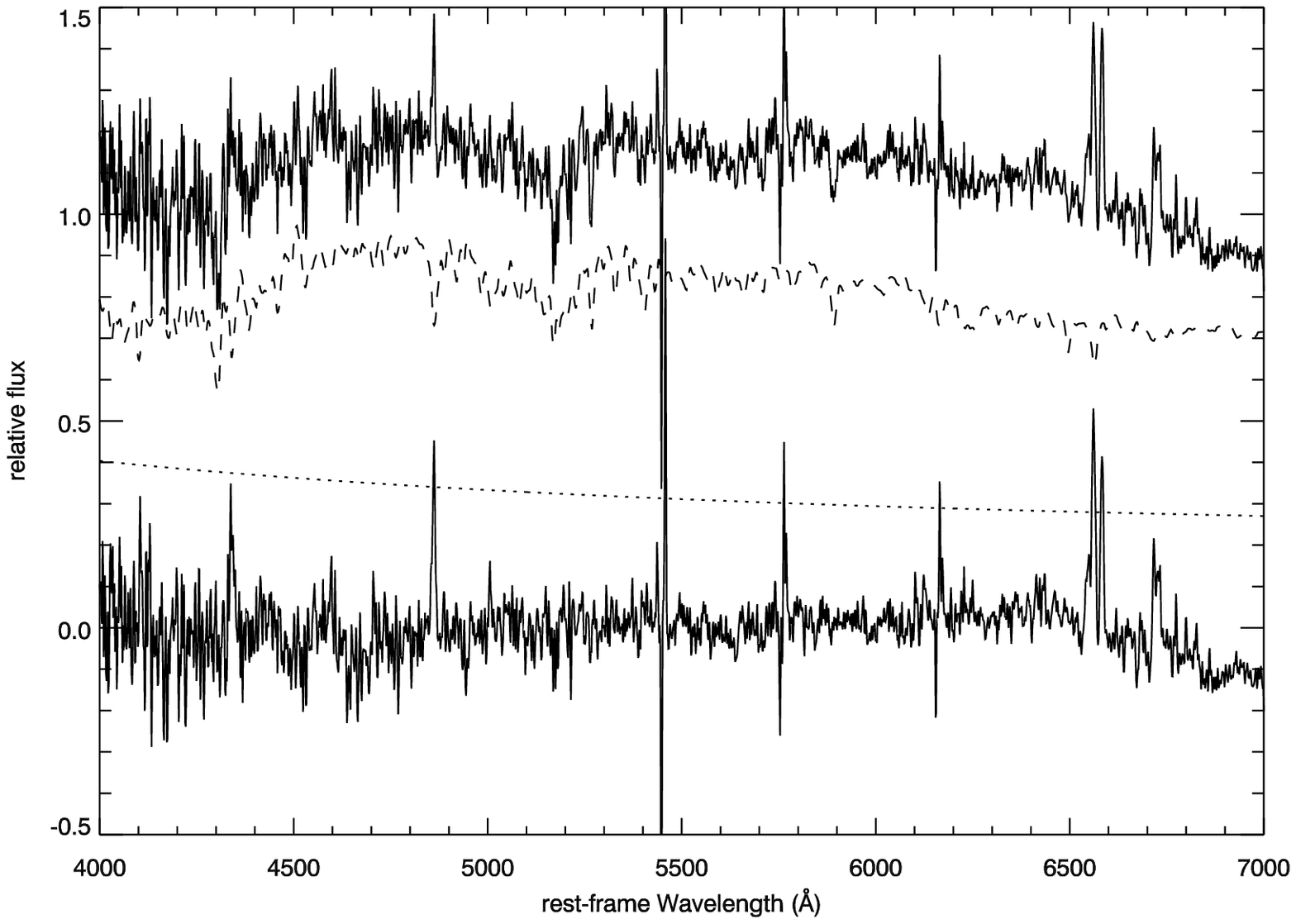}
&
\includegraphics[width=7cm]{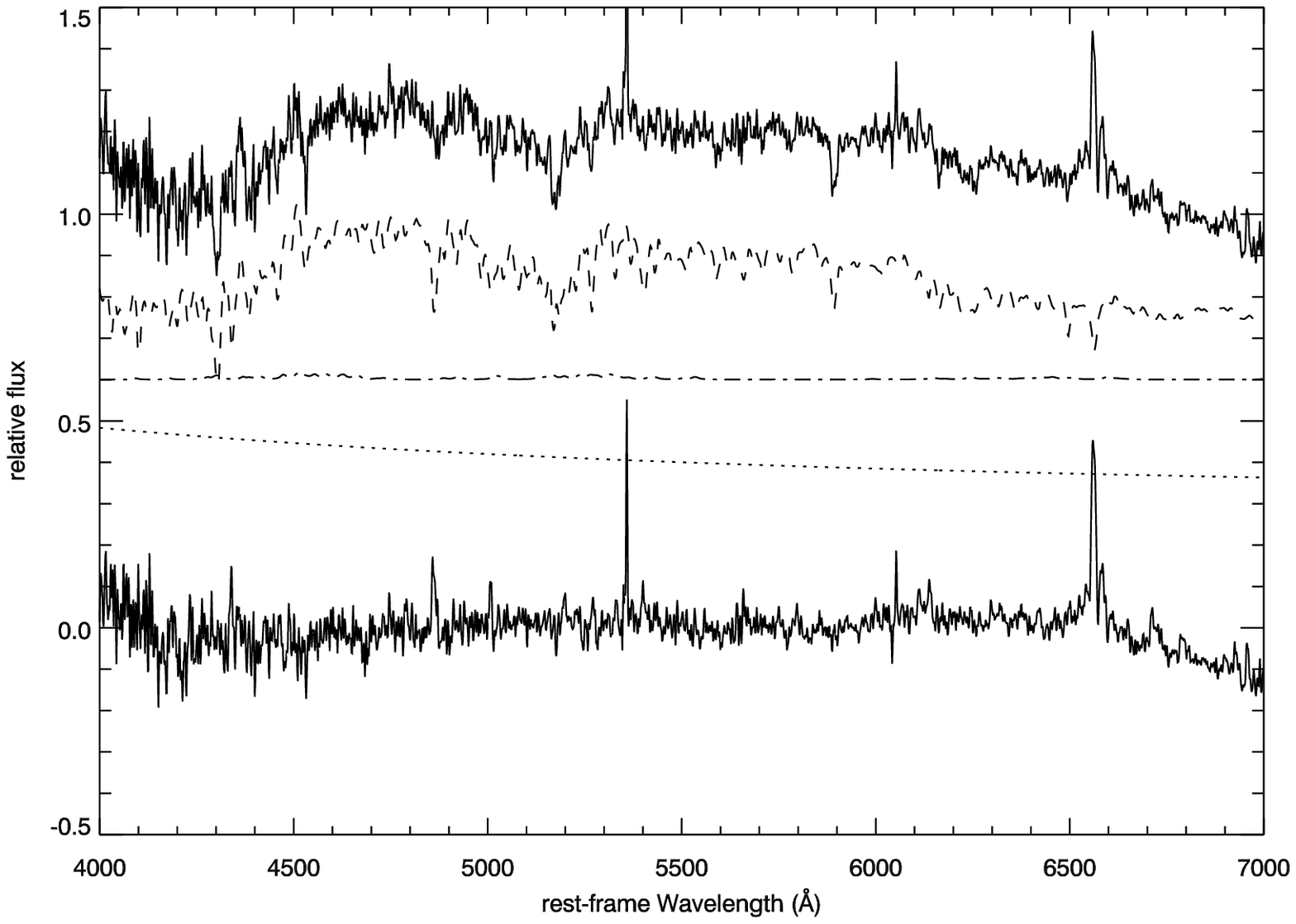} \\
\end{tabular}
\caption[]{Plots show observed HCT spectra on the top, with best fitting Gyr model showing the age of the
underlying stellar population as a dashed line. Fe\,{\sc i} and {\sc ii} template spectra taken from 
\cite{veron01} are plotted as dot-dashed. Dotted lines seen in UGC~1378, UGC~3968 and in UGC~4219 
are power-law continuum following \cite{mei09}. The bottom spectra drawn as thick solid 
lines show the subtracted spectrum. The x-axis represents the rest-frame wavelength. }%
\label{f:specfit1}%
\end{minipage}
\end{figure}

\begin{figure}%
\ContinuedFloat
\begin{minipage}{150mm}
\centering
\begin{tabular}{cc}
UGC 6614 & UGC 6754 \\
\includegraphics[width=7cm]{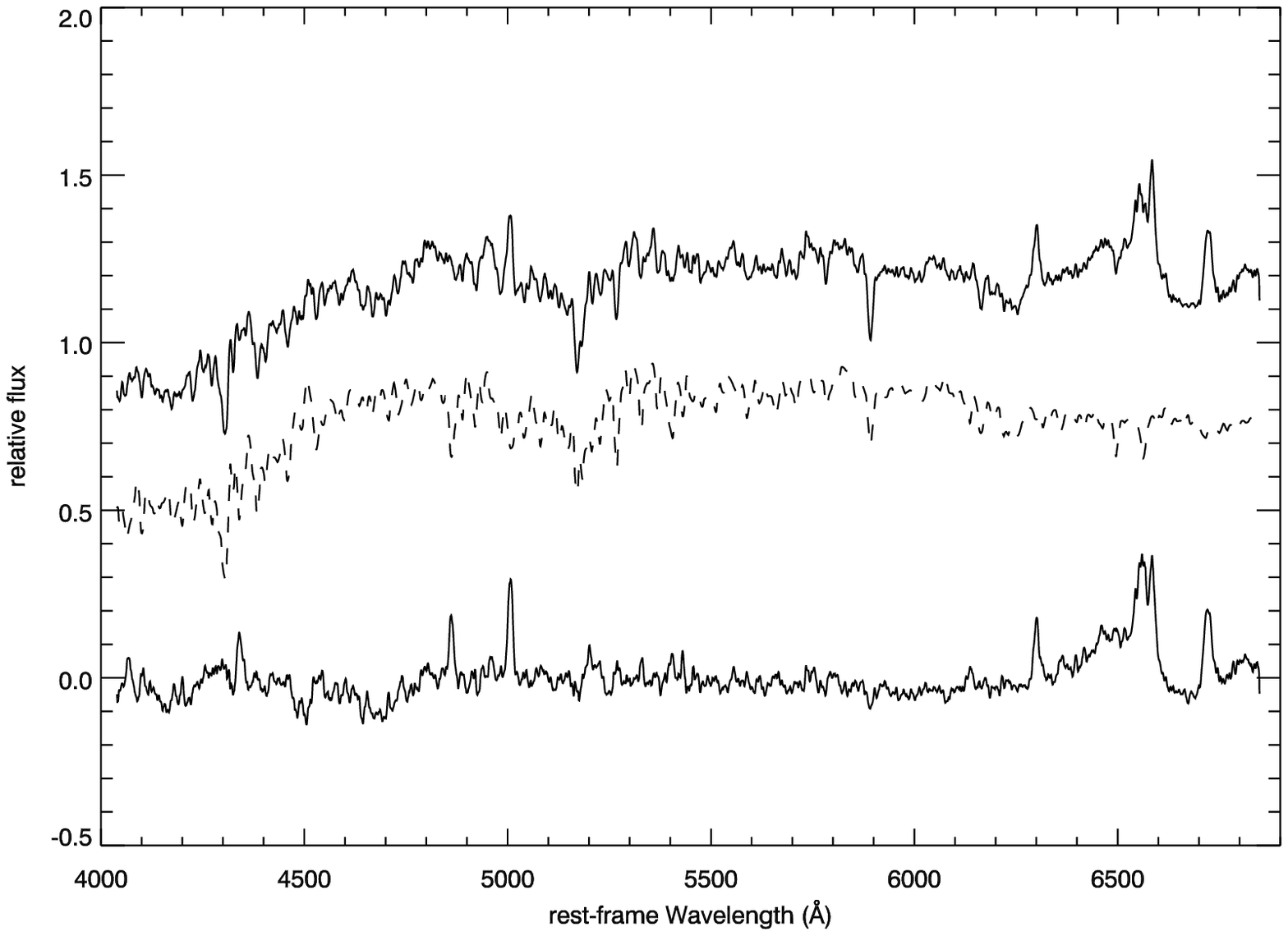}
&
\includegraphics[width=7cm]{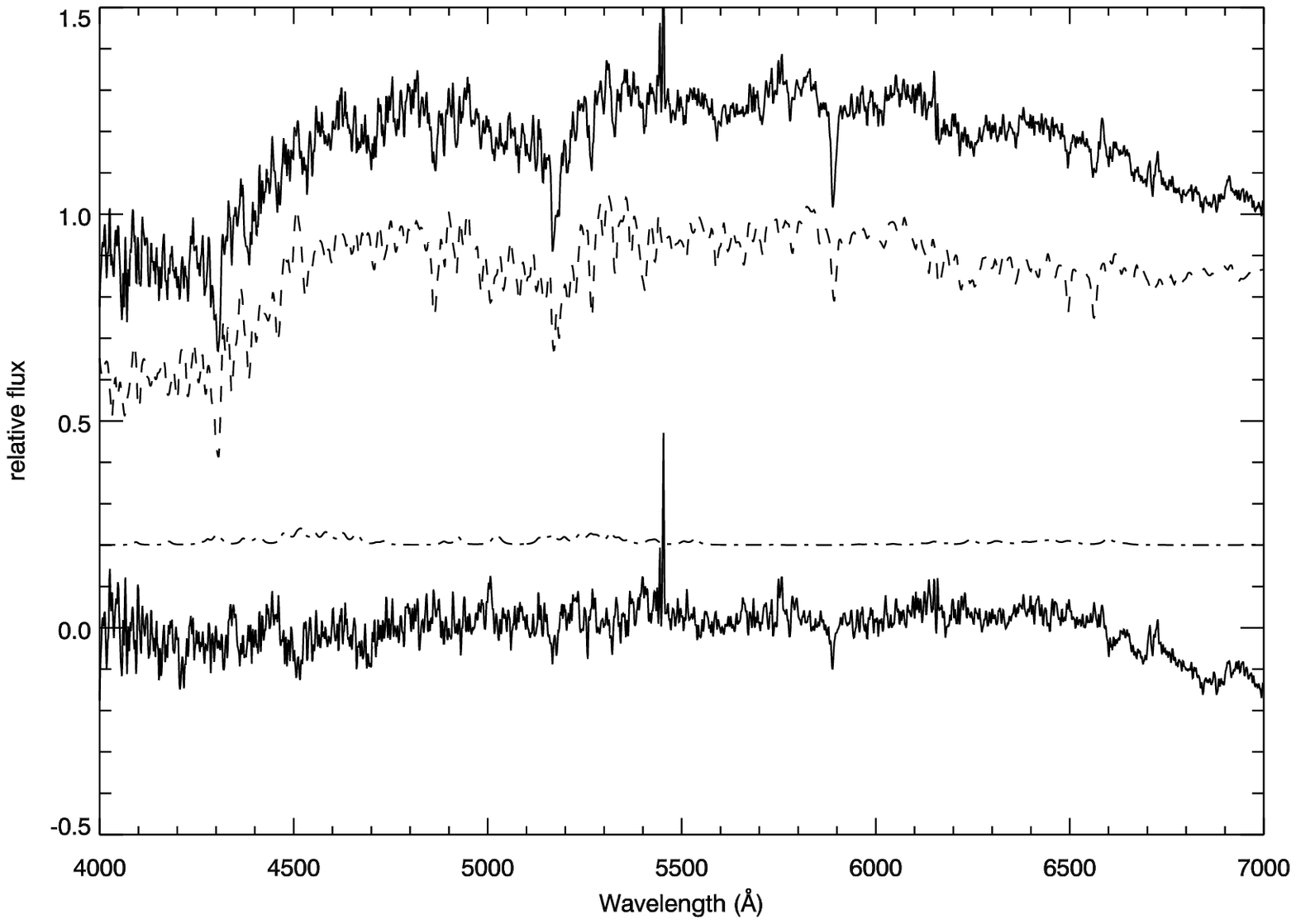} \\
UGC 6968 & UGC 7357 \\
\includegraphics[width=7cm]{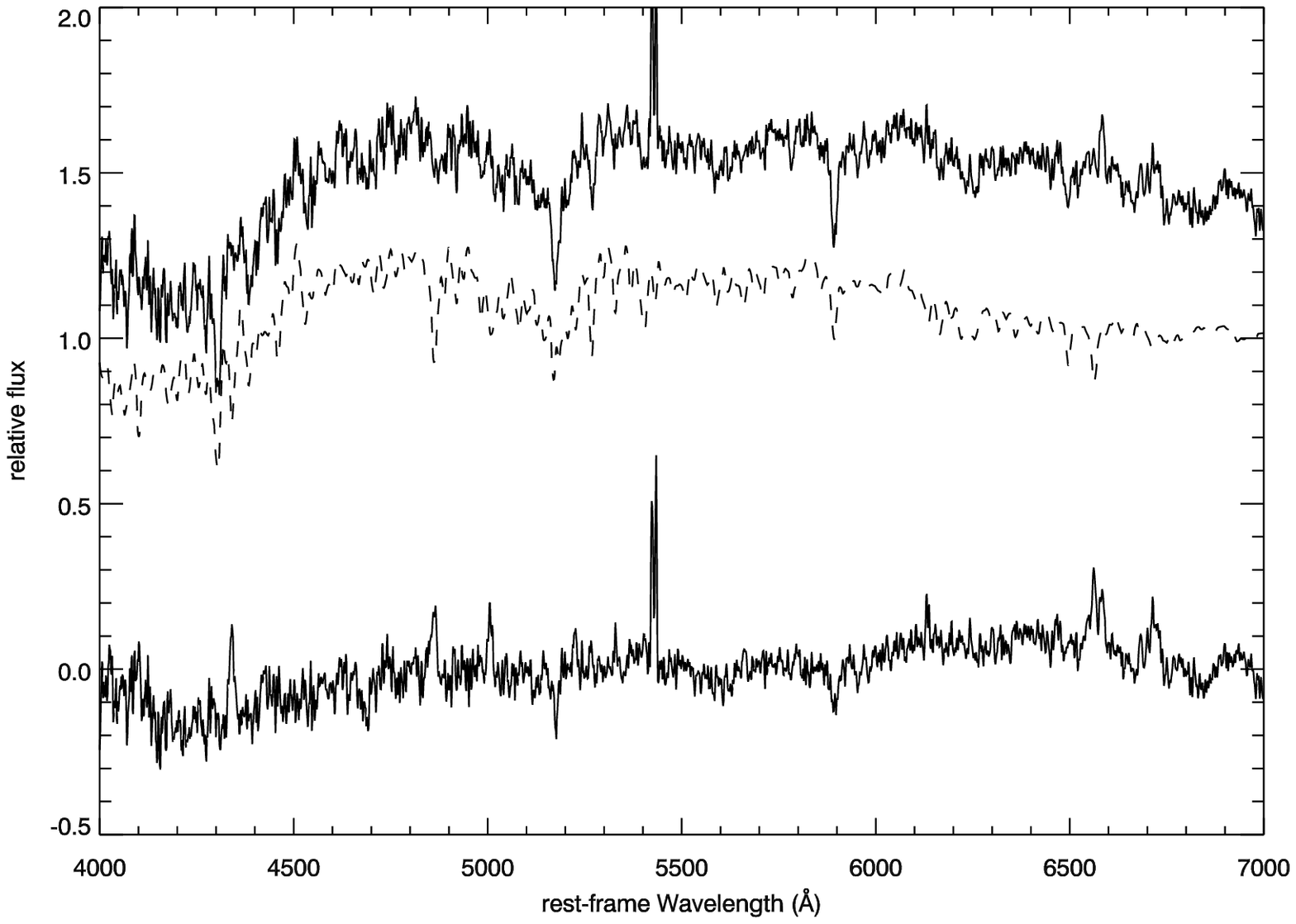}
&
\includegraphics[width=7cm]{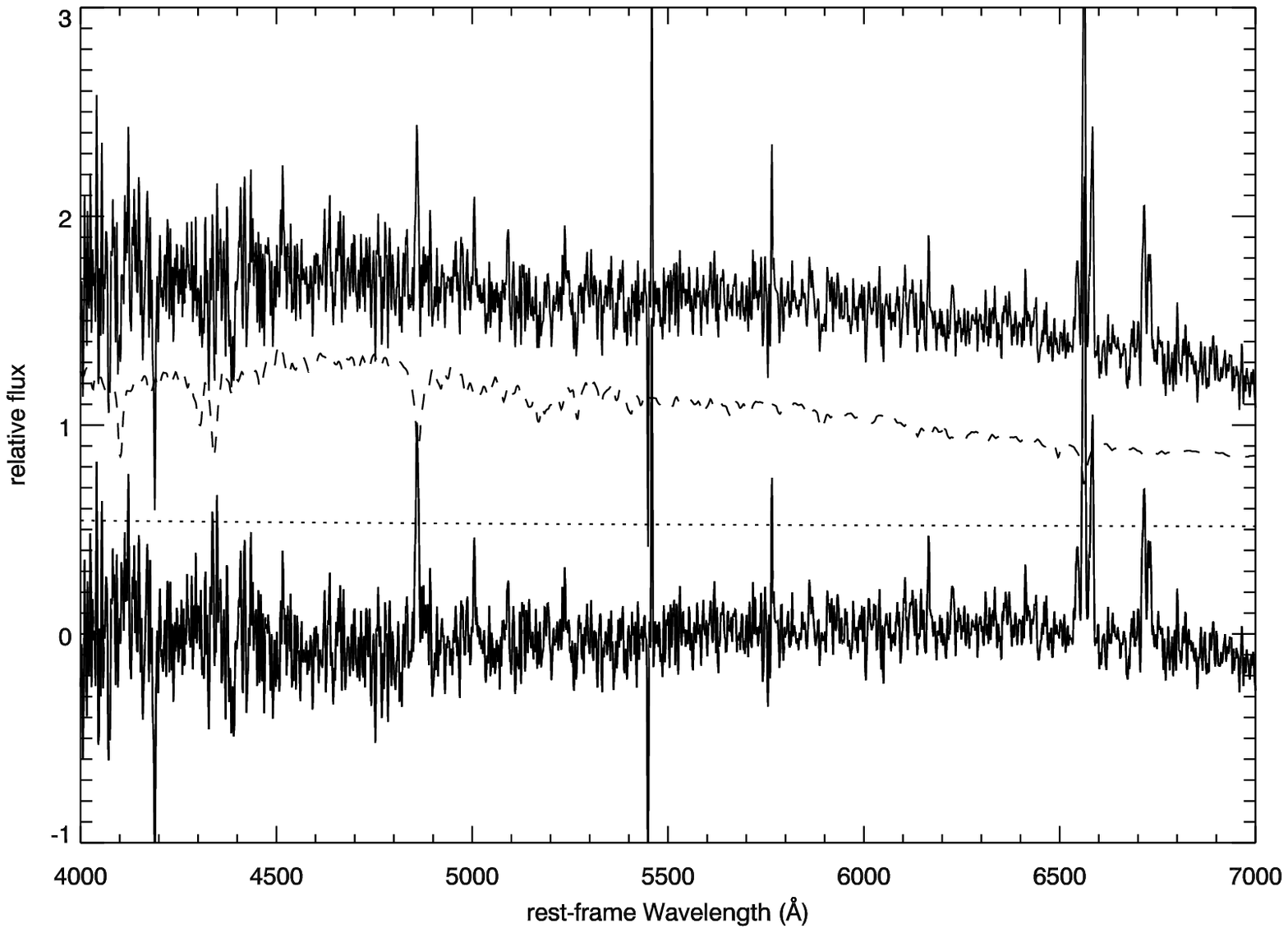} \\
LSBC~F568-6 & \\
\includegraphics[width=7cm]{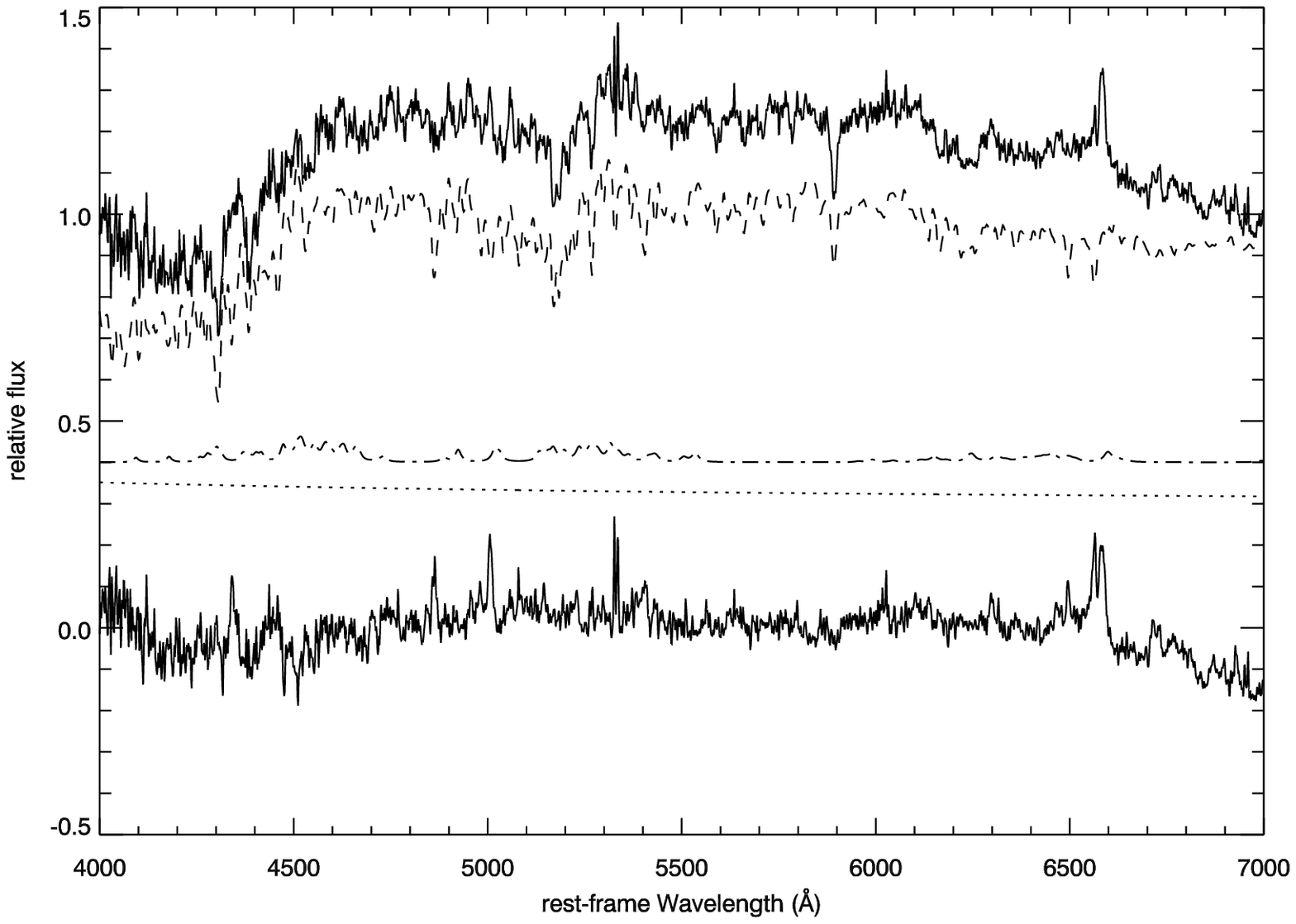}  &  \\
\end{tabular}
\caption[]{- \textit{Continued}. UGC 6614 and UGC 6968 are best fit using only the stellar spectra from 
Starburst99, while the spectra of UGC 7357 and F568-6 require additional power-law 
continuum shown as dotted lines in this figure. }%
\label{f:specfit2}%
\end{minipage}
\end{figure}

\begin{figure}%
\begin{minipage}{150mm}
 \centering
 \includegraphics[width=9cm]{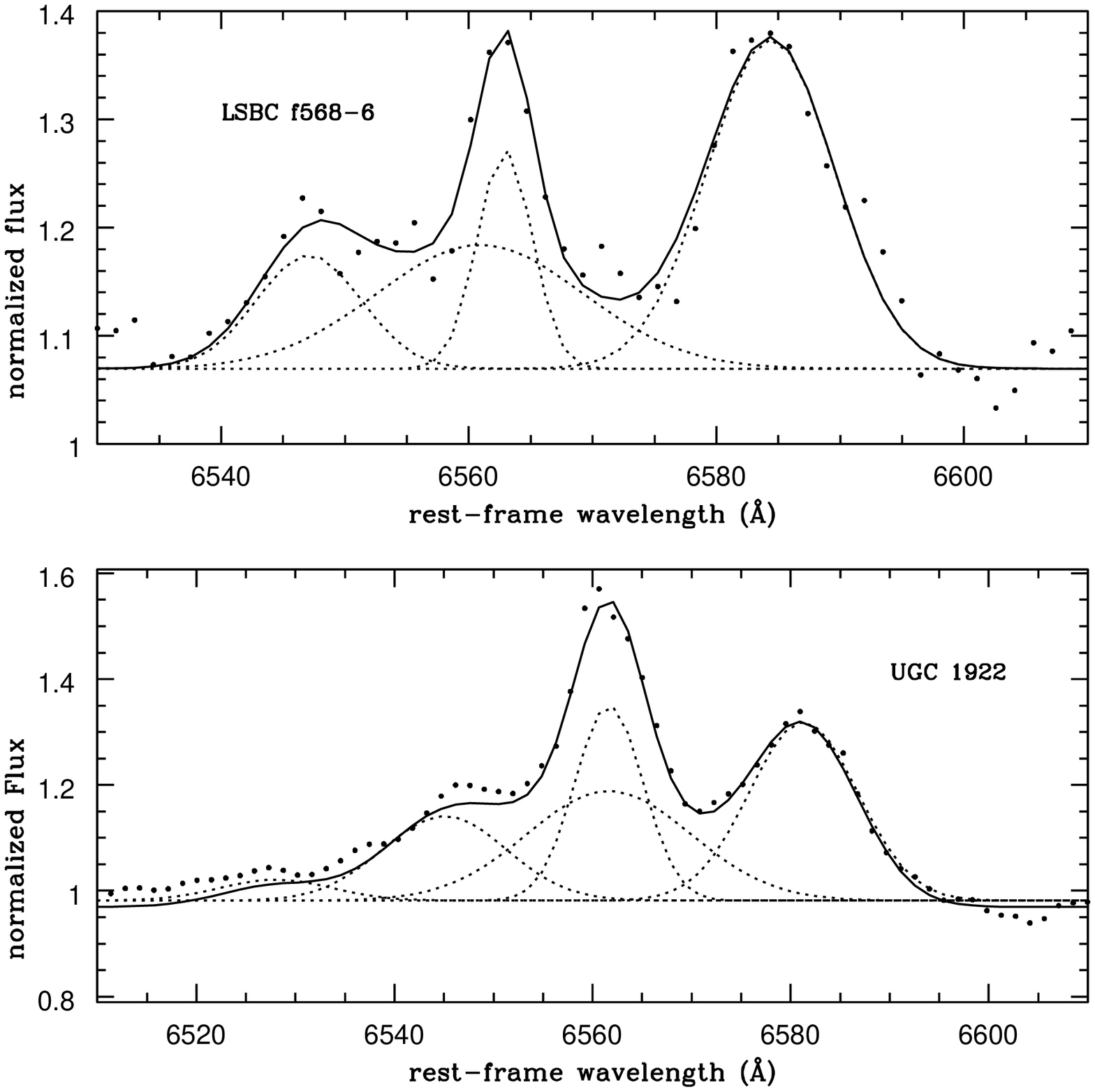}%
\caption{Fits to the broad and narrow emission line components for the galaxies F568-6 and UGC 1922 from SDSS. The 
points show the observed spectra after decomposition. The dotted lines show the individual gaussian fits to 
the lines and the solid line shows the combined fits for the region. The spectra used for estimating BH mass in F568-6 is 
the SDSS spectra while for UGC 1922, we have used HCT spectra as SDSS data for this galaxy is unavailable.}%
\label{f:lfit1}%
\end{minipage}
\end{figure}

\begin{figure}%
\ContinuedFloat
\begin{minipage}{150mm}
\centering
\includegraphics[width=9cm]{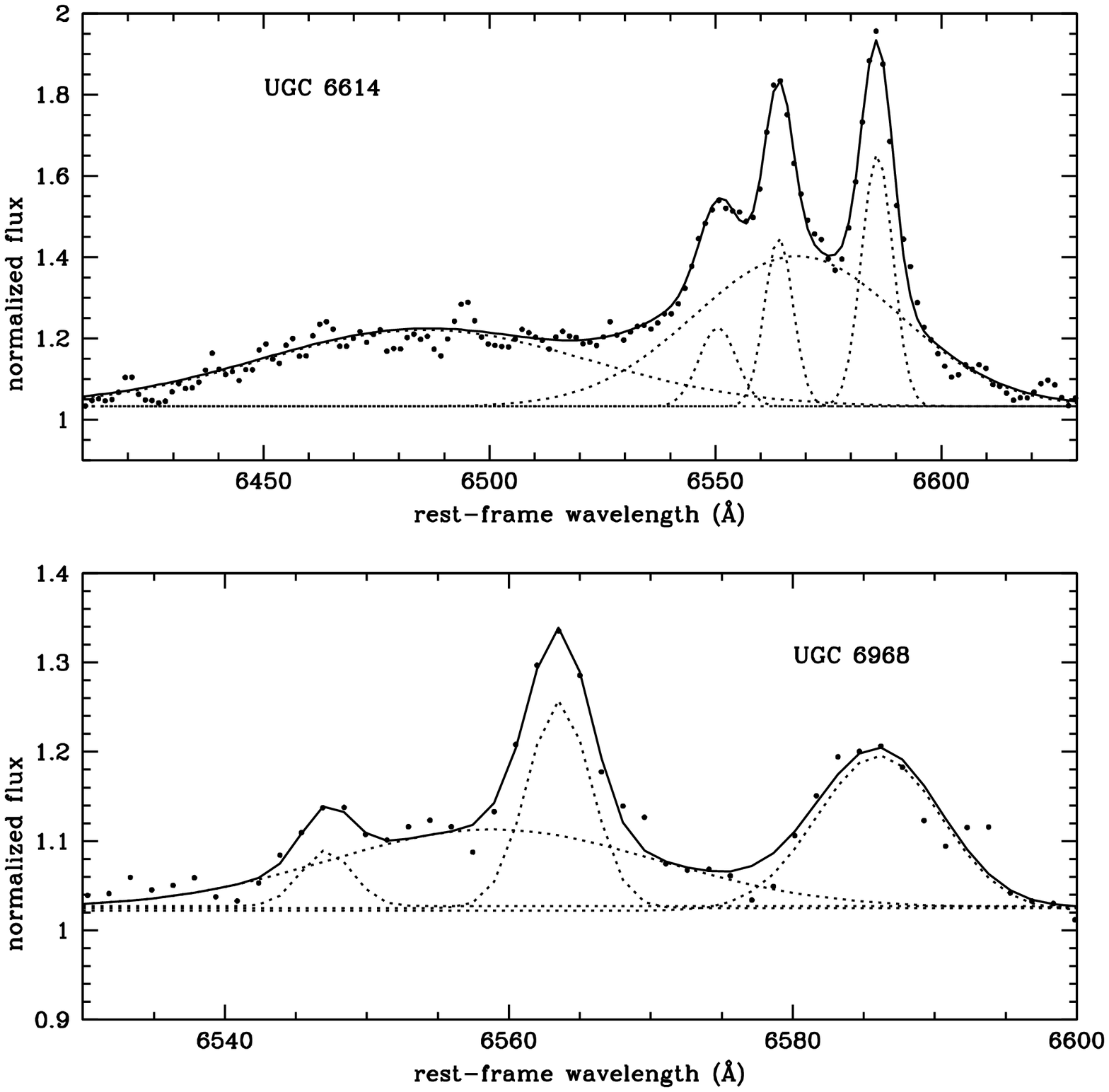}%
\caption[]{- \textit{Continued} using the SDSS spectra for the two galaxies UGC 6614 and UGC 6968.}%
\label{f:lfit2}%
\end{minipage}
\end{figure}

\begin{table}
\begin{minipage}{150mm}
\begin{center}
 \caption{The sample of LSB galaxies}  
\label{tabparam}
\begin{tabular}{ccccccccc}
 \hline\hline
Galaxy   & Other       & Galaxy     & Vel.$^a$& Distance$^b$ & Galaxy           & Galaxy      & Galaxy       & Absolute  \\
         & Name        & Type       & \kms & (Mpc)    & Coordinates  (J2000)   & inclination & size (arcmin) & Magnitude (M$_{\rm H}$) \\
\hline 
UGC 1378 & PGC~007247  & (R)SB(rs)a & 2935 & 43     &  01h56m19.2s +73d16m58s & 68.6$^{0}$ & 3.4  & -24.01 \\ 
UGC 1922$^c$ & PGC~009373  &  S     &10894 & 151    &  02h27m45.8s +28d12m33s & 29.6$^{0}$ & 2.1  & -25.25 \\
UGC 3968 & PGC~021636  & SB(r)c     & 6780 & 94     &  07h42m45.2s +66d15m30s & 37.8$^{0}$ & 1.4  & -23.55 \\
UGC 4219 & PGC~022766  & SA(rs)b    &12433 & 170    &  08h06m42.8s +39d05m25s & 44.7$^{0}$ & 1.1  & -24.37 \\
UGC 6614 & PGC~036122  & (R)SA(r)a  & 6352 & 86     &  11h39m14.9s +17d08m37s & 29.9$^{0}$ & 1.7  & -23.70 \\
UGC 6754$^d$ & PGC 036740 & SA(rs)b & 7025 & 96     &  11h46m47.2s +20d40m32s & 29.1$^{0}$ & 3.0  & -24.74 \\
UGC 6968 & PGC 037704  & S          & 8232 & 113    &  11h58m44.6s +28d17m22s & 44.7$^{0}$ & 2.8  & -24.04 \\
UGC 7357 & PGC 039640  & SAB(s)c    & 6682 & 91     &  12h19m13.4s +22d25m54s & 48$^{0}$   & 1.6  & -22.17 \\
F568-6   & Malin~2     & Sd/        &13830 & 189    &  10h39m52.5s +20d50m49s & 38$^{0}$   & 1.5  & -25.50 \\
\hline 
\end{tabular}
\end{center}
$^a$ Heliocentric velocities from NED
$^b$ Galactocentric distances from NED assuming $H_0=73$ \kms\/Mpc
$^c$ = IC 226
$^d$ = NGC 3883
\end{minipage}
\end{table}

\begin{table}
\begin{minipage}{150mm}
\begin{center}
 \caption{Observation details of long slit spectra obtained from {\bf HCT}.}
\label{tabobs}
\begin{tabular}{cccc}
 \hline\hline
{\bf Galaxy} & {\bf Date of Obs.} & {\bf Exptime}$^a$  \\
             &                    &  Gr7 / Gr8  in sec   \\
\hline 
UGC 1378 & 22-11-2006  & 3600 / 3600    \\
UGC 1922 & 22-12-2006  & 3600 / 3600    \\
UGC 3968 & 22-11-2006  & 1800 / 3600    \\
UGC 4219 & 19-02-2007  & 2400 / 2400    \\
UGC 6614 & 05-07-2006  & 3600 / 3600    \\
UGC 6754 & 19-02-2007  & 1800 / 2400    \\
UGC 6968 & 14-05-2007  & 1800 / 2400    \\
UGC 7357 & 19-02-2007  & 2400 / 1510    \\
F568-6   & 22-12-2006  & 3600 / 3600    \\
\hline 
\end{tabular}
\end{center}
$^a$ - Exposure times in the grism 7 (blue) and grism 8 (red).\\
\end{minipage}
\end{table}

\begin{landscape}
\begin{table}
\caption{Fluxes of emission lines in the units of $10^{-15}$ erg cm$^{-2}$ s$^{-1}$ obtained from {\bf HCT} spectra.}
\label{tabflux}
\begin{center}
\begin{tabular}{cccccccccc}
 \hline\hline
{\bf Galaxy} & {\bf [OII]} & {\bf \hb} & {\bf [OIII]} &  {\bf [OI]} & {\bf H$\alpha^a$} & \multicolumn{2}{c}{\bf [NII]} & \multicolumn{2}{c}{\bf [SII]} \\
             &  3727 \AA & 4861 \AA & 5007 \AA &  6300 \AA & $6563_{narrow}$ & 6548 \AA & 6584 \AA  & 6717 \AA & 6731 \AA  \\
\hline
UGC 1378 & ** & ** & $5.75\pm0.54$ & ** &  $5.30\pm0.81$ & ** & $8.20\pm0.87$ & $5.30\pm1.56$ & $4.74\pm1.25$ \\
 $^d$    &    &    & ($9.41\pm0.84$) &    &  ($10.12\pm1.19$) &  & ($10.16\pm0.98$) & ($13.07\pm3.12$) & ($12.72\pm3.03)$) \\
UGC 1922 & $31.01\pm0.07$ & $4.35\pm0.67$ & $3.32\pm0.56$ & $2.21\pm0.73$ & **$^b$ & $3.77\pm0.69$ & $8.29\pm0.59$ & $5.20\pm1.36$ & $3.56\pm0.89$\\
$^d$     & ($9.60\pm0.23$)&($9.47\pm1.08$) &($8.44\pm1.23$)& ($9.82\pm3.06$) & ** & ($15.58\pm1.88$) & ($18.72\pm3.08$) & ($11.67\pm2.60$) & ($9.93\pm2.95$) \\
UGC 3968 & ** & $2.08\pm0.13$ & ** & ** & $3.50\pm0.51$ & $1.81\pm0.52$ & $2.75\pm0.35$ & $0.91\pm0.40$ &  $1.45\pm0.59$\\
  $^d$   & ** & ($8.76\pm0.71$) & ** & ** & ($10.41\pm0.92$) & ($14.73\pm3.01$) & ($9.47\pm0.91$) & ($8.02\pm2.75$) & ($13.31\pm4.58$) \\   
UGC 4219 & ** & $1.10\pm0.25$ & ** & ** & $4.45\pm0.50$ & $0.58\pm0.35$ & $1.08\pm0.39$ & $0.69\pm0.45$ & $0.46\pm0.38$ \\
  $^d$   & ** & ($10.03\pm2.59$) & ** & ** & ($11.75\pm0.86$) & ($8.91\pm5.37$) & ($10.30\pm3.76$) & ($11.37\pm9.13$) & ($10.64\pm10.57$) \\
UGC 6614 & $7.89\pm0.09$ & $ 2.56\pm0.35$ & $5.12\pm0.31$ & $3.78\pm0.86$&  **$^b$  & $1.69\pm0.55$ & $5.58\pm0.40$ & \multicolumn{2}{c}{$5.46\pm0.36^c$} \\   
  $^d$   & ** & ($10.52\pm1.80$) & ($12.41\pm0.94$) & ($13.93\pm2.88$) & ($15.63\pm0.98$) & ($16.78\pm3.91$) & ($15.63\pm0.98$) & \multicolumn{2}{c}{($18.55\pm2.13$)}\\
UGC 6754 & ** & ** & $0.59\pm0.38$ & ** & $1.41\pm0.83$ & ** & $0.18\pm0.42$ & $0.54\pm0.33$ & $0.76\pm0.56$ \\
  $^d$   & ** & ** & ($8.59\pm6.63$) & ** & ($5.54\pm8.57$) & ** & ($5.83\pm4.84$) & ($7.34\pm4.46$) & ($11.15\pm8.04$) \\
UGC 6968 & ** & $4.52\pm0.62$ & $3.49\pm0.43$ & $1.14\pm0.41$ & **$^b$ & $0.40\pm0.53$ & $2.10\pm0.53$ &  $1.49\pm0.61$ & $2.14\pm1.09$ \\   
  $^d$   & ** & ($15.80\pm2.58$) & ($12.25\pm1.83$) & ($7.45\pm3.57$) & ** &  ** &** & ($6.78\pm2.60$) & ($15.06\pm7.56$) \\
UGC 7357 & ** & $2.03\pm0.08$ & $0.45\pm0.08$ & ** & $4.60\pm0.06$ & $1.16\pm0.09$ & $2.57\pm0.07$  & $2.31\pm0.16$ & $1.83\pm0.13$ \\
  $^d$   & ** & ($8.51\pm0.25$) & ($3.95\pm0.48$) & ** & ($8.21\pm0.16$) & ($10.37\pm0.78$) & ($9.85\pm0.36$) & ($10.25\pm0.36$) & ($11.12\pm0.79$) \\
F568-6   & *** & $1.66\pm0.26$ & $1.55\pm0.31$ & $0.49\pm0.39$ & **$^b$ & ** & $2.34\pm1.99$ & $0.79\pm0.61$ & $0.54\pm0.39$ \\
  $^d$ 	 & ** & ($12.57\pm2.23$) & ($10.17\pm1.56$) & ($9.65\pm2.33$) & ** & ** & ($15.31\pm1.31$) & ($11.56\pm7.78$) & ($9.77\pm7.51$) \\
\hline
\end{tabular}
\end{center}
$^a$ - H$\alpha$ fluxes for the objects which do not host AGN. \\
$^b$ - broad H$\alpha$ line is detected and hence separation of broad and narrow lines is done. These are 
deblended and given in the Table \ref{tabmbh}.\\
$^c$ - The telluric line $\lambda6870$ \AA \ exactly coincided with the [S\,{\sc ii}] emission. After the telluric 
line correction, the lines could not be resolved and hence the lines could not be deblended. Hence total 
[S\,{\sc ii}] flux is given here.\\
$^d$ - The values quoted inside the brackets, in the second row for each of the galaxies are FWHM
of each of the lines whose fluxes are given above. \\ 
\end{table}
\end{landscape}

\begin{table}
 \begin{minipage}{150mm}
 \begin{center}
\caption{\ha \ flux for narrow and broad components in erg cm$^{-2}$ s$^{-1}$, 
luminosity of broad \ha \ line, FWHM of the broad \ha \ line, $\sigma_*$ calculated using \ca2T \ lines
from SDSS spectra for three galaxies UGC 6614, UGC 6968 and F568-6. Due to unavailability of SDSS data for UGC 1922,
 HCT spectra is used for estimation of BH mass.}   
\label{tabmbh}
\begin{tabular}{ccccccccc}
 \hline\hline
{\bf Galaxy} &  \multicolumn{2}{c}{\bf H$\alpha^a$} & {\bf L$_{H\alpha}$} & {\bf fwhm$^b$} & {\bf $\sigma_*$} &  {\bf M$_{BH}$} \\
             &  Broad & Narrow & $\times 10^{40}$erg~s$^{-1}$ & km~s$^{-1}$ & km~s$^{-1}$  & $\times10^6$M$_\odot$ \\
\hline

UGC 1922 & $8.13\pm1.50$ & $4.79\pm1.93$ & $2.21\pm0.410$ & $855.7\pm140.7$ & ** & $0.39^{+0.18}_{-0.15}$ \\ \\
UGC 6614 & $32.77\pm3.62$ & $5.11\pm0.71$ & $2.91\pm0.020$ & $2456.5\pm167.3$ & $157.3\pm13.1$ & $3.89^{+1.21}_{-1.04}$ \\ \\
UGC 6968 & $3.54\pm1.68$ & $1.83\pm0.69$ & $0.54\pm0.001$ & $1244.7\pm742.8$ & $195.8\pm16.4$ & $0.45^{+0.81}_{-0.38}$ \\ \\
F568-6 & $2.01\pm0.84$ & $0.93\pm0.40$   & $0.85\pm0.020$ & $899.6\pm372.4$ & $209.1\pm18.0$ &  $0.29^{+0.32}_{-0.20}$ \\ \\
\hline 
\end{tabular}
\end{center}
$^a$- H$\alpha$ fluxes of the broad and narrow components in units of $10^{-15}$~erg~cm$^{-2}$~s$^{-1}$.\\
$^b$- FWHM of the broad line of H$\alpha$.\\
\end{minipage}
\end{table}
\end{document}